\documentclass[letterpaper,twocolumn,10pt]{article}
\usepackage{usenix2019_v3}

\usepackage{booktabs}

\usepackage{algorithm}
\usepackage{moreverb}
\usepackage{array}
\usepackage{multicol}
\usepackage{multirow}
\usepackage{caption}
\usepackage{amsthm}
\usepackage{enumitem}
\usepackage{makecell}
\usepackage[utf8]{inputenc}
\usepackage[english]{babel}
\usepackage{amsmath,amssymb}
\usepackage{tikz}
\usepackage{algpseudocode}
\usepackage{tcolorbox}
\definecolor{mygray}{gray}{.9}
\definecolor{mypink}{rgb}{.99,.91,.95}
\definecolor{myblue}{rgb}{.26,.45,.76}
\definecolor{mydarkorange}{rgb}{.237,.125,.49}
\definecolor{myorange}{rgb}{.255,.192,0}
\definecolor{mygreen}{rgb}{.112,.173,.71}
\definecolor{mylightblue}{rgb}{.180,.199,.231}
\usetikzlibrary{shapes,snakes}
\usetikzlibrary{positioning,arrows}
\usetikzlibrary{patterns}
\usetikzlibrary{fit}
\usepackage{pythonhighlight}
\newtheorem{theorem}{Theorem}

\usepackage[createShortEnv]{proof-at-the-end}

\makeatletter
\newenvironment{breakablealgorithm}
  {
   \begin{center}
     \refstepcounter{algorithm} 
     \hrule height.5pt depth0pt \kern1pt
     \renewcommand{\caption}[2][\relax]{
       {\raggedright\textbf{\ALG@name~\thealgorithm} ##2\par}%
       \ifx\relax##1\relax \addcontentsline{loa}{algorithm}{\protect\numberline{\thealgorithm}##2}%
       \else \addcontentsline{loa}{algorithm}{\protect\numberline{\thealgorithm}##1}%
       \fi
       \kern2pt\hrule\kern2pt
     }
  }{
     \kern2pt\hrule\relax
   \end{center}
  }
\makeatother

\usepackage{listings}
\usepackage{xcolor}

\definecolor{codegreen}{rgb}{0,0.6,0}
\definecolor{codegray}{rgb}{0.5,0.5,0.5}
\definecolor{codepurple}{rgb}{0.58,0,0.82}
\definecolor{backcolour}{rgb}{0.95,0.95,0.92}

\begin{document}

\title{Validated Strong Consensus Protocol for Asynchronous Vote-based Blockchains}

\author{
    Yibin Xu$^{1}$, Jianhua Shao$^{2}$, Tijs Slaats$^{1}$, Boris D{\"u}dder$^{1}$, Yongluan Zhou$^{1}$ \\
    $^{1}$\textit{Department of Computer Science, University of Copenhagen, Denmark} \\
    \{yx, boris.d, slaats, zhou\}@di.ku.dk \\
    $^{2}$\textit{School of Computer Science and Informatics, Cardiff University, UK} \\
    shaoj@cardiff.ac.uk
}
\maketitle

\begin{abstract}
Vote-based blockchains construct a state machine replication (SMR) system among participating nodes, using Byzantine Fault Tolerance (BFT) consensus protocols to transition from one state to another. Currently, they rely on either synchronous or partially synchronous networks with leader-based coordination or costly Asynchronous Common Subset (ACS) protocols in asynchronous settings, making them impractical for large-scale asynchronous applications.

To make Asynchronous SMR scalable, this paper proposes a \emph{validated strong} BFT consensus model that allows leader-based coordination in asynchronous settings. Our BFT consensus model offers the same level of tolerance as binary byzantine agreement but does not demand consistency among honest nodes before they vote. An SMR using our model allows nodes to operate in different,  tentative, but mutually exclusive states until they eventually converge on the same state. We propose an asynchronous BFT protocol for vote-based blockchains employing our consensus model to address several critical challenges: how to ensure that nodes eventually converge on the same state across voting rounds, how to assure that a blockchain will steadily progress through epochs while reaching consensus for previous epochs, and how to maintain robust byzantine fault tolerance.

Our protocol greatly reduces message complexity and is the first one to achieve linear view changes without relying on threshold signatures. We prove that an asynchronous blockchain built on our protocol can operate with the \emph{same} simplicity and efficiency as partially synchronous blockchains built on, e.g. HotStuff-2. This facilitates deploying asynchronous blockchains across large-scale networks.
\end{abstract}

\section{Introduction}
Byzantine Fault Tolerance (BFT) consensus protocols form the backbone of State Machine Replication (SMR) systems. These protocols enable honest nodes to agree on a shared value and then make state transitions \cite{lamport1982the,pease1980reaching}, despite the presence of potential adversarial nodes. Their recent implementations are in vote-based blockchains.

Traditionally, vote-based blockchains use binary agreement, where nodes either vote to accept or reject a block, typically proposed by a periodically elected leader. They require the existence of \emph{at most} one block for each voting instance. The honest nodes must make the same decision for the block in order to maintain consistency. They are commonly used in blockchain implementations, such as Tendermint~\cite{buchman2016tendermint}, Algorand~\cite{gilad2017algorand}, and HotStuff~\cite{yin2019hotstuff}. However, in the asynchronous communication model, where no time-bound assumptions are made, it becomes impossible to discern whether a leader is controlled by the adversary (i.e., chooses not to share its block) or just being slow (i.e., other nodes do not receive its block in time). Setting a time-out to replace a silent leader is not a solution as it contradicts the communication model. Thus, these protocols cannot guarantee liveness in consensus derivation in asynchronous settings as they may not terminate.

Asynchronous Common Subset (ACS) based blockchain protocols \cite{abraham2019asymptotically, miller2016honey, gao2022dumbo} address the liveness issue by assembling a block for advancing a blockchain using transaction sets submitted by at least \(N-f\) nodes in an \(N\)-node system with \(f\) potential Byzantine nodes. Liveness is ensured as honest messages will eventually be delivered. ACS-based protocols usually require \(O(N^3)\) messages due to using Multi-valued Validated Byzantine Agreement (MVBA) protocols or concurrently using multiple instances of Asynchronous Binary Agreement (ABA) protocols. Moreover, they can lead to transaction overlaps and greatly increase the protocol's cost, even with a small number of distinct transactions. 

This paper proposes, to the best of our knowledge, the first BFT consensus protocol that successfully addresses the liveness issue in asynchronous networks using leader-and-vote-based designs. We introduce a new consensus model, termed \textit{Validated Strong} BFT Consensus, which extends the classical Strong BFT consensus~\cite{10.1145/872035.872066,neiger1994distributed}. This model enables nodes to agree on a single value from a set of $m \geq 2$ possible input values. In the context of blockchain, each input value corresponds to the decision of accepting a specific block from a set of $m$ candidate blocks. This contrasts with traditional binary Byzantine agreements, which limit nodes to deciding between only two options---either accepting or rejecting the block proposed by the current leader.

We implement our BFT consensus model for blockchains and show that it allows multiple mutually exclusive blocks to be voted on in each voting round until one is consensually accepted. A new leader is elected periodically and proposes a new block to compete with the previous ones until one is consensually accepted. The periodically elected new leader does not replace the old leaders even if their blocks have not been delivered in time; the consensually accepted block can be proposed by either the new or the old leaders. So, it does not violate the assumption of asynchronous networks.

Moreover, honest nodes do not need to vote for the same block in each voting round. As such, an honest node does not need to wait until it receives \(N-f\) acknowledgments for the same view number before voting for a block in that view. This design allows a leader node to broadcast blocks without requiring \(N-f\) nodes to recognize it as the leader first, making our design the first to use only a single message and \(O(N)\) message complexity for a view change, even without threshold signatures\footnote{HotStuff~\cite{yin2019hotstuff} uses threshold signatures to achieve \(O(N)\) message complexity for view changes. Message complexity~\cite{cachin2000random} is the (expected) number of messages sent to the network. E.g., if a node sends a message of its own to all the \(N\) nodes, the message complexity would be \(O(N)\).}.

Importantly, our Validated Strong BFT consensus model preserves the fault tolerance of the binary Byzantine agreement, allowing up to $f\leq \lfloor(N-1)/3\rfloor$ byzantine faulty nodes in asynchronous networks. This is a departure from the classical Strong BFT model, which imposes a stricter fault tolerance of $f \leq \lfloor (N-1)/(m+1) \rfloor$, where the number of faulty nodes is inversely proportional to the number of input values.



The salient features of our protocol include:

\noindent \textbf{Gradual Consensus Establishment:} Nodes may start from distinct tentative states and converge via multiple voting rounds until a consensus state is reached. It does not require honest nodes to have the same view before they vote.

\noindent \textbf{Instant-Runoff Voting:} It eliminates less supported blocks round-by-round until one block remains. This not only gradually aligns honest nodes but also gradually limits voting options for Byzantine nodes. When only one block remains, a silent vote 
    implies acquiescence rather than disagreement.

\noindent  \textbf{Efficient Progression in Asynchronous Settings:} By nesting several consensus derivation instances, leader nodes can propose new blocks of greater block heights regardless of whether the preceding block has been accepted or not. The fast nodes can advance to new blockchain epochs after voting in the current epoch without waiting for sufficient votes from others, allowing the system to progress steadily despite the presence of slower nodes.

\noindent \textbf{Reduced Complexity:} Our protocol significantly decreases the message complexity. It achieves an \(O(m+\theta N)\)  message count and requires a message complexity of \(O(mN+\theta N^2)\) if each node sends its vote directly to every other node, where \(m\) represents the number of blocks and $\theta$ is the number of voting rounds used. If implemented in a manner similar to HotStuff, where the leader collects votes, the total message is reduced to at most \(O(mN+\theta N)\). In a standard asynchronous network with arbitrary but finite network delays, our protocol can use as small as two rounds of voting in the best case before the consensus is established . This is consistent with binary agreement protocol HotStuff-2~\cite{malkhi2023hotstuff}, which operates only in synchronous or partially synchronous environments. Our protocol requires an amortized \(\log_2(mN)\) voting rounds,  only in unrealistic scenarios where the adversary deterministically controls the message delays of all nodes. 






\noindent \textbf{Message Forwarding Enabled:} The message complexity for ACS-based and DAG-based protocols \cite{keidar2021all,danezis2022narwhal,spiegelman2022bullshark} reaches $O(N^3)$, making communication costly if every message is signed by the sender. So, many protocols do not sign the messages but use fully meshed and authenticated channels without message forwarding. Our reduced complexity practically allows signing each message, supporting data re-transmission and broadcast via Gossip protocols. This makes our protocol suitable for large-scale deployment.

In summary, we make the following contributions:

1). We introduce a validated strong BFT consensus model and show that an asynchronous leader-and-vote-based SMR is possible with this novel consensus model. 

2). We present a scalable asynchronous blockchain protocol based on a validated strong BFT consensus model, reducing  message count to \(O(N)\). 

3). We show that our protocol 
    outperforms ACS-based protocols Dumbo2 and Dumbo-NG across various factors, demonstrating its suitability for large-scale applications.

\section{Problem Domain}
\label{new2}
Here, we give the system model assumptions, the definition and challenges of Validated Strong BFT consensus.
\subsection{Assumptions on System Model}
Our objective is to construct a replicated service ensuring continuous consensus on input values and establishing a chain of consensus values among $N$ distributed nodes. The following assumptions are necessary for our system:

\noindent \textbf{General Assumptions.} We assume the availability of hash functions, digital signatures, and a public-key infrastructure (PKI). The system operates with a preset $N$ number of nodes.

\noindent \textbf{Strong Adaptive Adversary.} A adversary comprises $f \leq \lfloor (N-1)/3 \rfloor$ Byzantine nodes, capable of deviating from the standard protocol through various means. 
Once the adversary attempts to corrupt a node, the node is immediately corrupted. Moreover, messages sent from the node before corruption but not yet delivered can be erased (or clawed back) by the adversary.

\noindent \textbf{Security Assumptions.} We assume that at least $N-f$ nodes are not adversarial and will conform to the standard protocol of operations outlined in this paper.

\noindent \textbf{Network Assumptions.} Our framework adopts the standard asynchronous communication assumption, where a message sent from one honest node to another eventually reaches its destination, albeit subject to arbitrary, unknown, yet finite communication delays. We support any network topology as long as this assumption is held.


\subsection{Validated Strong BFT Consensus}\label{VSC}
In an $N$-node system using the Validated Strong BFT consensus model, for each consensus-establishing instance, there are $m \leq N$ distinct input values proposed by different nodes. Suppose that we have a global polynomial-time computable predicate $Q(v, \pi)$ that can be used to validate an input value $v$ with a proof method $\pi$. Among the $m$ input values $v_1,v_2 \ldots v_m$, at least one input value $v_j$ will be validated as true by $Q$. The Validated Strong BFT consensus model essentially implements this predicate and relies on three essential properties for achieving consensus:

1). \textbf{Termination}: All honest nodes eventually commit.

2). \textbf{Agreement}: All honest nodes commit to the same and only one input value.

3). \textbf{Validated Strong Validity}: If an honest node commits to a value $v_i$, then $Q(v_i, \pi)$ must hold true and $v_i$ is the input value of some nodes.

Validated Strong Validity differs from Strong Validity in that it involves validation of input values:

\textbf{Strong Validity \cite{10.1145/872035.872066,neiger1994distributed} :} If an honest node commits to an input value $v_i$, then $v_i$ is the input value of some honest nodes.

This difference is significant in terms of security, as illustrated by the following examples. Suppose we have a self-driving car capable of receiving $m$ independent instructions from $N \geq m$ centers on what to do next. We know that among them, a maximum of $f$ centers may not be reliable (i.e., could give wrong instructions), but we do not know which ones. Hence, to ensure a consensual decision is safe in the end, we must ignore any instruction that is echoed by fewer than $f$ centers. This is because the system is unable to validate if a received instruction is correct or not. In other words, strong validity lacks the ability of predicate $Q(v, \pi)$ in operation. Strong BFT consensus can, therefore, tolerate $f \leq \lfloor (N-1) / m \rfloor$ and $f \leq \lfloor (N-1) / (m+1) \rfloor$ for synchronous and asynchronous settings, respectively.

In contrast, a blockchain validates a block $B$ (if $B$ contains transactions that contradict previous blocks) by reconciling $B$ and the previous blocks of $B$, which form $\pi_B$. Honest nodes abstain from committing $B$ if $Q(B,\pi_B)$ is false, regardless of how many nodes have voted for it. As honest nodes will only vote for a correct block, and if a protocol can assist all honest nodes to eventually converge to the same $B'$ before committing, where $Q(B',\pi_{B'})$ holds true, then validated strong BFT consensus can tolerate $f\leq \lfloor (N-1)/2\rfloor$ and $f\leq \lfloor (N-1)/3\rfloor$ for synchronous and asynchronous settings respectively, a clear security advantage over strong BFT consensus.

\noindent\textbf{Comparison with multi-valued validated byzantine agreement.}
Multi-valued validated byzantine agreement~\cite{cachin2001secure} (used in ACS-based protocols) commits multiple input values as long as they are validated.
Validated strong BFT consensus, in contrast, commits only \emph{one} input value among multiple ones.

\begin{table}[h!]
\centering
\caption{Byzantine Agreement (BA) Primitives}
\label{tab:consensus_models}\footnotesize
\begin{tabular}{@{}p{2.3cm}p{3cm}p{3cm}@{}}
\toprule
\textbf{Primitive} & \textbf{Input Output} & \textbf{Asynchronous Byzantine Fault Tolerance} \\ \midrule

\textbf{Binary BA\ \ (Atomic Broadcast)} & \textbf{Input}: One value\newline \textbf{Output}: Commit/abort   &  $f\leq \lfloor (N-1)/3\rfloor$ \\ \midrule

\textbf{Multi-valued Validated BA} & \textbf{Input}: Multiple values \newline \textbf{Output}: All validated  &    $f\leq \lfloor (N-1)/3\rfloor$ \\ \midrule

\textbf{Strong BFT Consensus} & \textbf{Input: } Multiple values \newline \textbf{Output}: One &$f \leq \lfloor (N-1) / (m+1) \rfloor$ \\\midrule
\textbf{Validated Strong BFT Consensus} & \textbf{Input: } Multiple values \newline \textbf{Output}: One validated & $f\leq \lfloor (N-1)/3\rfloor$ \\
 \bottomrule
\end{tabular}
\end{table}
\vspace{-.1cm}
\hspace{-.4cm}\fbox{%
  \parbox{.5\textwidth}{%
    \textbf{Voting Complexity Comparison in a Nutshell:} 
    Nodes in Binary BA, Strong BFT consensus and Validated Strong BFT consensus model makes \emph{only one} vote per voting round, until the protocol terminated. In contrast, MVBA requires nodes to find \emph{all} validated input values and vote on each and everyone of them for their validity, resulting in much higher voting complexity.
  }%
}

\vspace{-.2cm}
\section{Validated Strong Consensus Protocol for Blockchain with Byzantine Fault Tolerance}\label{new3}

\subsection{Key Elements of the Protocol}

To help understand our proposal, we first highlight four key aspects of our protocol that depart from existing asynchronous protocols for achieving strong BFT consensus:

1). \textbf{Periodic Leader-based Block Proposal and Allowing Unknown $m$:} The classical protocol for strong BFT consensus (see Appx.~\ref{strong}) doesn't specify the origin of the $m$ input values. Implementing it in an asynchronous leader-based blockchain would require $m \geq N-f$ nodes to propose blocks. In contrast, our protocol uses a single periodically rotating leader node to propose a block in each time window. However, a block can arrive at any time due to arbitrary but finite message delays. We guarantee the correctness of the consensus protocol even with inconsistencies of blocks among nodes and new blocks being discovered during voting rounds.

2). \textbf{Increased Adversarial Tolerance:} The classical protocol tolerates $f\leq \lfloor (N-1)/(m+1) \rfloor$ adversarial nodes, $m>2$. Our protocol allows $f\leq \lfloor(N-1)/3 \rfloor$. 

3). \textbf{Forcing Verdict Unification via Voting Rules:} A vote is structurally linked to $N-f$ votes of the last round. A partial voting history can be recursively obtained via a vote graph starting from this vote. This structure enables verification of voting rules compliance. Specifically, any recipient of a vote can unambiguously obtain this partial voting history and determine whether the vote was validly cast for a block. In other words, the voter must justify why it voted for a particular block --- namely, that it was the most supported block according to the voting history. This design ensures that nodes gradually converge on a single block. Over time, the Byzantine nodes lose the ability to justify voting for alternative blocks.

4). \textbf{Concurrent Consensus Derivation for Multiple Instances:} In contrast to existing BFT protocols for blockchains, which run only one instance at a time, our protocol nests and concurrently executes multiple instances. Specifically, each blockchain height \(k \geq 0\) starts a new instance \(k\). The voting process is structured such that a vote cast at height \(k\) corresponds to a round 0 vote for instance \(k\), a round 1 vote for instance \(k-1\), and so on, up to a round \(k\) vote for instance 0.



\subsection{Definitions}\label{Definitions}

\noindent \textbf{Consensus Chain.} The consensus chain refers to the sequence of valid blocks maintained by all nodes. A valid block does not contain incorrect transactions or transactions that contradict those in the previous blocks of the consensus chain. When a block $B_{k}$ extends a block $B_{k-1}$, there is a directed link from $B_{k}$ to $B_{k-1}$. 

\noindent \textbf{Vote Chain.} The vote chain for node $i$ consists of successive votes, where each vote of node $i$ at height $k \geq 1$ has a directed link to node $i$'s vote at height $k-1$. 

\noindent \textbf{Height.} Consensus or vote height $k \geq 0$ indicates the specific position in the consensus or vote chain, preceded by $k$ blocks or votes.

\noindent \textbf{Vote ($V^{(k)}_i$).} This is the vote cast by node $i$ for a block at consensus height $k$. For $k \geq 1$, $V^{(k)}_i$ is a tuple consisting of the hash value of a block that node $i$ votes for at height $k$ and the hash values of $N-f$ votes from height $k-1$.


\noindent \textbf{Connected Vote Graph ($G_i^{(k)}$).} This is a directed graph built progressively as vote height increases. $G_i^{(k)}$ can be traversed recursively from the vote $V^{(k)}_i$ to votes in vote height 0 following the links attached to it. In other words, $G_i^{(k)}$ contains all the votes reachable from $V^{(k)}_i$.

\noindent \textbf{Vote View ($\operatorname{VV}_i^{(k)}$).} 
This is an array of integers, where $\operatorname{VV}_i^{(k)}[B]$ is defined for a block $B$ of height less than $k$. 
$\operatorname{VV}_i^{(k)}[B]$ contains the sum of the votes in $G_i^{(k)}$, which were cast on $B$. 
Note that $\operatorname{VV}_i^{(k)}$ does not count $V^{(k)}_i$.

\noindent \textbf{Vote Count ($\operatorname{VC}_i^{(k)}$).} 
This is an array of integers, where $\operatorname{VC}_i^{(k)}[B]$ sums the votes cast on block $B$ and any blocks up until consensus height $k-1$ stemming from $B$, using the data from $\operatorname{VV}_i^{(k)}$.

\subsection{Consensus Chain Propagation}\label{outputting}
\begin{figure}[h!]
    \centering
    \includegraphics[width=0.5\textwidth]{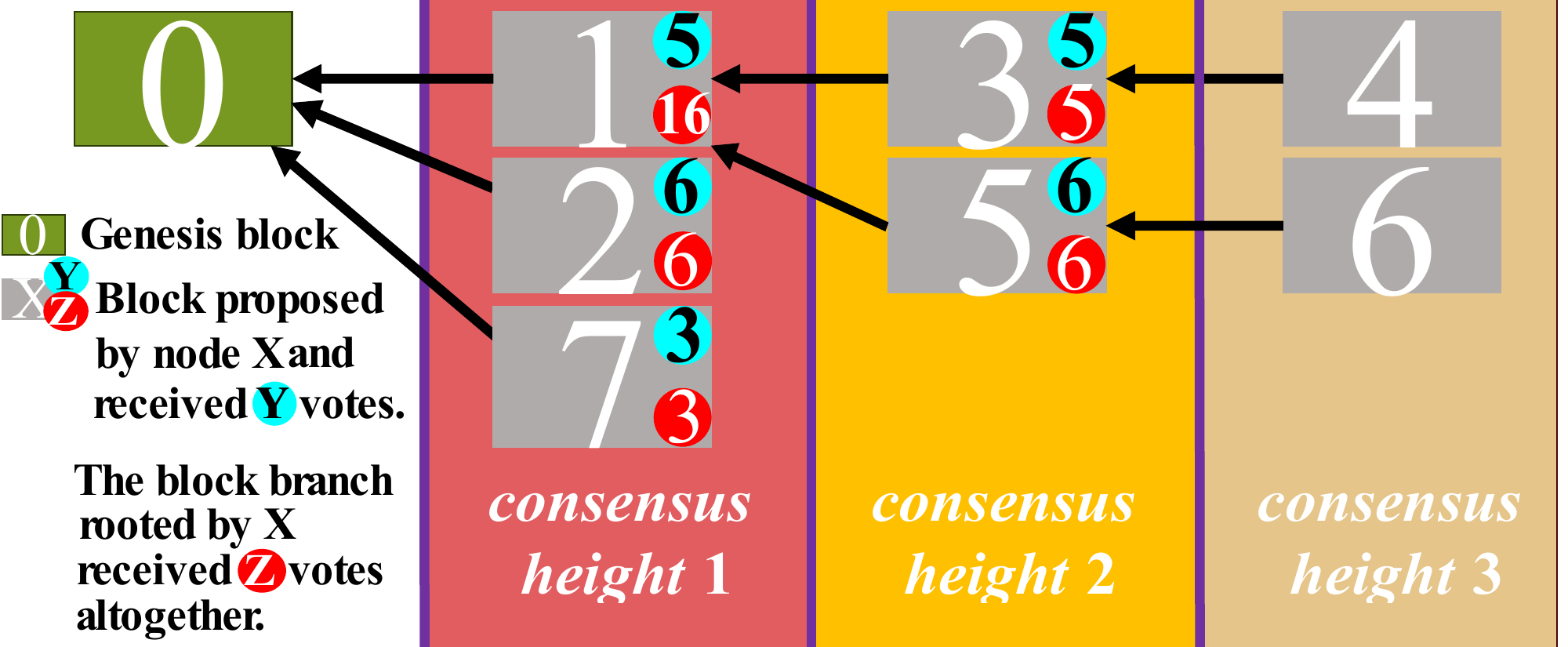}
\vspace{-.7cm}\caption{An example of consensus chain propagation: Node 1 generated Block 1, stemming from Block 0. Block 2 did not extend Block 1; instead, it extended Block 0. Block 3 extended Block 1, etc. This figure also illustrates the vote view $\operatorname{VV}_j^{(k)}$ and the vote count $\operatorname{VC}_j^{(k)}$, for $k=3$. The numbers in the blue and red circles on Block X are $\operatorname{VV}_j^{(3)}[X]$ and $\operatorname{VC}_j^{(3)}[X]$ respectively. $G_j^{(3)}$ contains votes until height 2 to the knowledge of node $j$.}
    \label{fig:outputblock}\vspace{-.5cm}
\end{figure}

For simplicity of explanation, we assume all $N$ nodes are indexed and have access to a local clock\footnote{For example, there can be a pre-set agreement for nodes to start the system together at 00:00 UTC on a day.}, denoted by $T$, which is \emph{not} required to align perfectly among the nodes. Node 0 generates the Genesis block at $T=0$, which is consensually accepted by all nodes.
At the beginning of each predefined period $\Delta BI$, the node with index $i=\lfloor\frac{T}{\Delta BI}\rfloor \mod N$ proposes a block to extend the consensus chain known to node $i$ \footnote{Branches (forkings) may exist, resulting in multiple consensus chains. Because of arbitrary but finite network delays, these consensus chains may or may not be known to node $i$.}.  
As nodes are not assumed to be aligned precisely according to $T$, the adversary could effectively propose a block at any time for any branch. Further discussion on this design see Appx.~\ref{C3}. Fig.~\ref{fig:outputblock} shows an example of consensus chain propagation. All blocks stem from the Genesis block and form a tree structure. 

\subsection{Voting}\label{lba}
All votes at height 0 are for the Genesis block. An honest node $i$ casts a vote at vote height $k>0$ only upon receipt of $N-f$ votes of height $k-1$ \footnote{Note that if some of the $N-f$ votes are labeled as non-compliant votes by the node, this node will wait until it receives $N-f$ compliant votes. Sec.~\ref{4.4.1} discusses when a vote will be labeled non-compliant.}, which can be used to construct $G_{i}^{(k)}$ and derive both $\operatorname{VV}_{i}^{(k)}$ and $\operatorname{VC}_{i}^{(k)}$ \footnote{Note that $G_{i}^{(k)}$ may contain double-voting votes, which need to be deducted before deriving $\operatorname{VC}_{i}^{(k)}$. See the filtering rule in Sec.~\ref{4.4.1}.}.

\noindent \textbf{Honest model.} Honest nodes follow these voting rules:
\begin{enumerate}[leftmargin=0.4cm, itemindent=-0cm,itemsep=2pt,topsep=0pt,parsep=0pt]
    \item Honest nodes vote only once per vote height.\label{p1}
    \item  \label{p2} At vote height 1, an honest node $i$ can vote for any valid block of consensus height 1 stemming from the Genesis block. At vote height $k > 1$, it votes for a valid block $B_k$ that satisfies the following conditions:
    \begin{enumerate}[leftmargin=0.5cm, itemindent=-0cm, itemsep=2pt,topsep=0pt,parsep=0pt]
        \item \label{p21} Let $B_{k'}$ be the block at consensus height $k'$ that $B_k$ stems from, where $0\leq k' < k$, and $B_0$ be the Genesis block. $\forall k'\in \{1,...,k-1\}$, $B_{k'}$ must have the highest vote count among blocks at consensus height $k'$ that stem from $B_{k'-1}$ according to $\operatorname{VC}_i^{(k)}$.  
        \item   If multiple blocks $B_k$ meet the condition~, node $i$ votes for one of them at vote height $k$ at random.
    \end{enumerate}
  
\end{enumerate}

Alg.~\ref{a1} shows the pseudocode for the voting process used by any honest node $i$.

\begin{algorithm}[h!]
\caption{Voting Process}
\begin{algorithmic}[1]\small
\Function{CastVote}{$i$, $currentHeight$}
    \If{$currentHeight \geq 1$}
        \State Wait for $N-f$ compliant votes of $currentHeight-1$.
         \State Generate $V_{i}^{(currentHeight)}$ that links to $N-f$ compliant votes of $currentHeight-1$. Build $G_{i}^{(currentHeight)}$ and then calculate $\operatorname{VC}_{i}^{(currentHeight)}$.
        \State Cast $V_{i}^{(currentHeight)}$ on a block of $currentHeight$ from the most supported branch (per criteria.~\ref{p2} of the honest model, according to $\operatorname{VC}_{i}^{(currentHeight)}$).
    \Else
    \State  Cast $V_{i}^{(0)}$ on the Genesis block.
    \EndIf

\EndFunction
\end{algorithmic}
\label{a1}
\end{algorithm}

\noindent \textbf{Vote reasoning.} When receiving \(V_j^{(k)}\) from any node $j$, any node can acquire the blocks known to node $j$ and their associated votes via \(G_j^{(k)}\). Fig.~\ref{fig:outputblock} shows an example with six blocks up to height 3 and the information derived from \(V_j^{(3)}\). The numbers in the blue and red circles on block \(X\) are $\operatorname{VV}_j^{(3)}[X]$ and $\operatorname{VC}_j^{(3)}[X]$ respectively. No information about block 4 and block 6 are indicated in \(V_j^{(3)}\) as \(G_j^{(3)}\) only contains votes up until height 2. One cannot infer if node $j$ knows any block of height 3 base on \(V_j^{(3)}\). But, \(V_j^{(3)}\) must be cast for block 6 as it is in the branch rooted by block 5, which has the highest vote count among blocks of height 2 stemming from block 1. Block 1 has the highest vote count among blocks of height 1 that stems from the Genesis block.

\noindent \textit{Remarks:} The reason for using \(\operatorname{VC}\), which sums up the votes of a block's branch as the vote count of this block, is that once this block is consensually accepted, later consensually accepted blocks must come from its branch. Thus, the goal is to decide on which branch (rooted by a block) is most supported, not merely the most supported block.

\subsubsection{Vote validation}\label{4.4.1}
Honest nodes only accept votes that were cast following the honest model. We define the adversarial model in order to validate votes. A Byzantine node must exhibit at least one of the two behaviours:

(C1) Vote more than once at the same vote height (the double-voting attack).

(C2) Fail to vote for blocks according to criteria~\ref{p2} of the honest model.

When receiving a vote $V_j^{(k)}$, a node first constructs $G^{(k)}_j$ using the hash pointers in the vote. It then filters out any double-voting votes from $G_j^{(k)}$. Afterwards, it derives $\operatorname{VV}_j^{(k)}$ and $\operatorname{VC}_j^{(k)}$ and checks if $V_j^{(k)}$ is a compliant vote. 

\begin{figure}[h!] \centering \vspace{-.25cm}\includegraphics[width=0.35\textwidth]{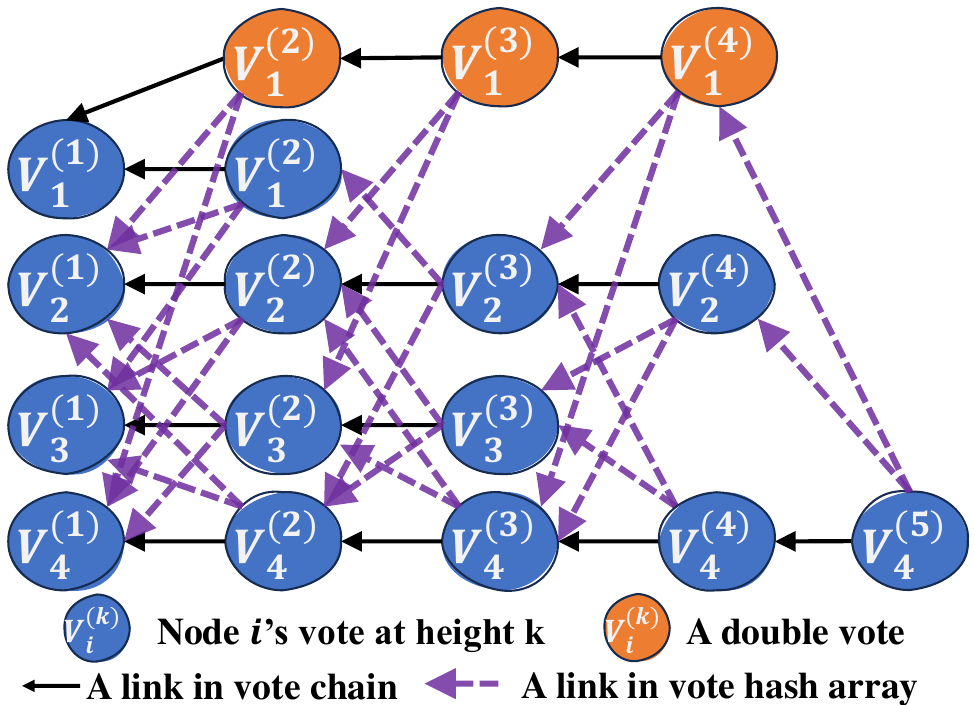}\vspace{-.25cm}\caption{A visualization of $G_4^{(5)}$ } \label{fig:my_label32432}\end{figure}

\noindent \textbf{Filter out double voting votes:}  Double-voting is detected by checking if two or more votes from the same node at the same height exist in $G^{(k)}_j$. If double-voting by any node $j'$ is detected in $G^{(k)}_j$, only the votes prior to the first fork of the vote chain of node $j'$ is counted in $\operatorname{VV}_j^{(k)}$. Fig.~\ref{fig:my_label32432} depicts a scenario of a four-node system. When a node receives vote $V^{(5)}_4$, it reconstructs $G^{(5)}_4$ as shown in the figure. From this graph, $V_4^{(5)}$ is linked to $V_1^{(4)}$, $V_2^{(4)}$  and its own $V_4^{(4)}$. Analysis of the paths in the graph reveals a double-voting for $V_1^{(2)}$, identified through distinct paths $V_4^{(5)}\rightarrow V_1^{(4)} \rightarrow V_1^{(3)} \rightarrow V_1^{(2)}$ and $V_4^{(5)}\rightarrow V_2^{(4)} \rightarrow V_2^{(3)}\rightarrow V_1^{(2)}$, etc. Any node receives $V^{(5)}_4$ can reconstruct an identical $G^{(5)}_4$.
Following the detection of double-voting votes, all subsequent votes related to the anomaly are removed from $\operatorname{VV}_4^{(5)}$ and will not be counted in $\operatorname{VC}_4^{(5)}$.

\noindent In practice, the first fork of one's vote chain can be detected in linear time using reverse breadth-first searching on $G^{(k)}_j$.

\noindent \textbf{Verify the compliance of voting rules:} If a vote is not classified as a compliant vote, it is classified as a non-compliant vote. $V_j^{(k)}$ is a compliant vote if it satisfies all the following points: (1) $V_j^{(k)}$ was cast for a block at consensus height $k$; (2) $G_j^{(k)}$ does not contain non-compliant votes; (3) $V_j^{(k)}$ was cast for a block $B_k$, which is in the branch rooted by $B_{k'}$ of consensus height $k'\in [1,k-1]$. $B_{k'=1}$ originated from the genesis block. Each other block $B'_{k'}$ of consensus height $k'$, which is in the same branch as $B_{k'}$ (they both stem from $B_{k'-1}$ of consensus height $k'-1$), fulfills   
\begin{equation}
\operatorname{VC}_j^{(k)}[B_{k'}]\geq \operatorname{VC}_j^{(k)}[B'_{k'}]\label{acc1}
\end{equation}

Alg.~\ref{a2} shows the pseudocode for vote validation. Lem.~\ref{k1} shows that \emph{all} nodes must vote in the most supported branch suggested by the Vote Count of their votes. Further justifications for compliance checking see Appx.~\ref{C2}.

\begin{algorithm}[h!]\small
\caption{Vote Validation}
\begin{algorithmic}[1]
\Function{ReceiveVote}{$V^{(k)}_j$}
    \State $G_j^{(k)} \gets$ \Call{ConstructConnectedGraph}{$V^{(k)}_j$}
    \State Filter out double voting votes on $G_j^{(k)}$
    \State $\operatorname{VV}_j^{(k)} \gets$ \Call{DeriveVoteView}{$G_j^{(k)}$}
        \State $\operatorname{VC}_j^{(k)} \gets$ \Call{DeriveVoteCount}{$\operatorname{VV}_j^{(k)}$}
        \State Verify if $V^{(k)}_j$ follows voting rules using data in $\operatorname{VC}_j^{(k)}$ and $G_j^{(k)}$
    \If{ $V^{(k)}_j$ is a non-compliant vote}
        \State Ignore $V^{(k)}_j$
    \EndIf
\EndFunction
\end{algorithmic}
\label{a2}
\end{algorithm}

\begin{theoremEnd}[end, restate]{Lemma}
\textbf{(Compliance to the Voting Rule)} Suppose that a node $j$ cast a $V_j^{(k)}$,  $V_j^{(k)}$ will be linked by any compliant vote at height $k+1$ only if $V_j^{(k)}$ is a compliant vote.
\label{k1}
\end{theoremEnd}
\begin{proofEnd}
\textbf{Assumption: }We assume the existence of a robust adversarial model and compliance-checking mechanisms as described in Sec.~\ref{4.4.1}, particularly focusing on filtering non-compliant votes.

\textit{Step 1: Assume for Contradiction}

Assume that the lemma is false. That is, assume there exists an adversarial node \( j \) whose vote \( V_j^{(k)} \) for a block stemming from \( B_{k'} \) at height \( k \) is linked by an honest node at height \( k+1 \), even though \( B_{k'} \) is not stem from the most supported branch according to \( \operatorname{VC}_j^{(k)} \).

\textit{Step 2: Adversarial Vote Analysis}

Under the voting rules, a compliant vote is only linked to a vote if it satisfies the criteria outlined for a compliant vote. According to the criteria, a node \( i \) will only consider \( V_j^{(k)} \) if:
- The block \( B_{k'} \) voted on by \( j \) has the highest vote count among all blocks at height \( k' \) of the same branch and this branch stems from the most supported branch, according to \( \operatorname{VC}_j^{(k)} \). 

\textit{Step 3: Invoking the Voting Rule and Mechanisms}

Given that \( V_j^{(k)} \) was linked by an honest node \( i \) at height \( k+1 \), it implies that:

\begin{itemize}
    \item \( V_j^{(k)} \) adhered to the honest voting criteria~\ref{p2}, contradicting the assumption that \( B_{k'} \) was not the root of the most supported branch stemming from the most supported branch according to \( \operatorname{VC}_j^{(k)} \) .
    \item Therefore, if \( B_{k'} \) was indeed not the root of the most supported branch stemming from the most supported branch based on \( \operatorname{VC}_j^{(k)} \), the honest nodes would have disregarded \( V_j^{(k)} \) due to the enforcement of compliance checks \( \operatorname{VC}_j^{(k)}[B_{k'}] < \operatorname{VC}_j^{(k)}[B'_{k'}] \) for any other block \( B'_{k'} \) with higher support.
\end{itemize}

\textit{Step 4: Security Reduction}

By reduction, if \( V_j^{(k)} \) were accepted and linked by an honest node without \( B_{k'} \) being the root of the most supported branch stemming from the most supported branch, it would imply that the voting mechanism failed to correctly enforce its rules or detect non-compliance. This failure would represent a breakdown in the integrity of the voting process, directly contradicting the protocol's assumption of reliable, compliant vote counting and validation mechanisms.

\textbf{Conclusion}: Given the system's design to filter non-compliant votes and enforce the honest model, it is contradictory to assume a scenario where an adversarial vote for a less supported branch gets linked by an honest node. So, by contradiction, the lemma holds under the protocol's operational assumptions and security model. Thus, a Byzantine node must vote for the most supported branch to their local knowledge to have their vote considered by others.
\end{proofEnd}
\subsection{Reaching Consensus}\label{reaching}
The Genesis block is consensually accepted. Node $i$ considers all honest nodes to have reached a consensual acceptance on a block $B_{k'>0}$ at the vote height $k>k'$ when both of the conditions are met: (I) The preceding block of $B_{k'}$ is consensually accepted. (II) There exists a group of nodes of size at least $N-f$, denoted by $NG$, that meets \begin{equation}
\begin{aligned} \forall j\in NG:\quad \quad \quad 
\operatorname{VC}_{j}^{(k)}[B_{k'}] - t  > \\\operatorname{VC}_{j}^{(k)}[B'_{k'}(j)]+\operatorname{VC}_{j}^{(k)}[U_{k}]-\operatorname{VC}_{j}^{(k)}[U_{k'}]-1
                     \label{acc7}\end{aligned}
        \end{equation}  where 
        \begin{itemize}[leftmargin=0.2cm, itemindent=-0cm,itemsep=2pt,topsep=0pt,parsep=0pt]
        \item $V_j^{(k)}$ is a compliant vote.
            \item $t=(k-k') \times f$  
         is the maximum \textit{potential} double-voting votes between vote height $k'$ to $k-1$.
         \item $B'_{k'}(j)$ is the second most supported block of the consensus height $k'$ according to $\operatorname{VC}_j^{(k)}$. If there is no block other than $B_{k'}$, then $\operatorname{VC}_{j}^{(k)}[B'_{k'}(j)]=0$.
         \item $\operatorname{VC}_{j}^{(k)}[U_{k'}]$ is the number of votes from height $1$ to $k'-1$ (inclusive) that are yet to be received by node $j$ according to $G_{j}^{(k)}$~\footnote{There are $N\times (k'-1)$ votes from height $1$ to $k'-1$, one can locate all the received votes from $G_{j}^{(k)}$, then calculate the missing vote number.} 
        \end{itemize}

Determining if a block $B_{k'}$ at consensus height \(k'\) is accepted at vote height \(k\) involves the following steps:

1). Check the received votes of height \(k\) for compliance.

2). Check if conditions (I) and (II) are met.

Alg.~\ref{Main} describes the pseudocode for reaching consensus.

\begin{breakablealgorithm}
\captionof{algorithm}{Reaching Consensus on a Block}\small
    \label{Main}
\begin{algorithmic}[1]
\Function{IsConsensusReached}{$B_{k'}$, $k'$, $k$}
          \If{$k'=0$}
          \State \textbf{return} $B_{k'}==genesis\ block$
          \EndIf
  \State $supportingNodes \gets$ \Call{GetSupportingNodes}{$B_{k'}$, $k$}
    \If{$\text{size of } supportingNodes \geq N-f$}
        \State $consensusCount \gets 0$
        \ForAll{$j \in supportingNodes$}
            \State $voteForB_{k'} \gets$ \Call{VoteCount}{$B_{k'}$, $\operatorname{VV}_j^{(k)}$}
            \State $voteForB'_{k'} \gets$ \Call{VoteCount}{$B'_{k'}(j)$, $\operatorname{VV}_j^{(k)}$}
            \State $missingVotesAtK' \gets$ \Call{MissingVotesCount}{$k'$, $G_j^{(k)}$}
            \State $missingVotesAtK \gets$ \Call{MissingVotesCount}{$k$, $G_j^{(k)}$}
            \State $potentialDoubleVotes \gets (k - k') \times f$
            \If{$voteForB_{k'} - potentialDoubleVotes > voteForB'_{k'} + missingVotesAtK - missingVotesAtK' - 1$}
                \State $consensusCount \gets consensusCount + 1$
            \EndIf
        \EndFor 
   \State $precedingBlock \gets \text{PrecedingBlock}(B)$ 
        \If{$consensusCount \geq N-f \And$ \Call{IsConsensusReached}{$precedingBlock, k'-1, k$}}
            \State \textbf{return} \textbf{true}
        \EndIf
    \EndIf
     \State \textbf{return} \textbf{false}
\EndFunction

\Function{GetSupportingNodes}{$B_{k'}$, $k$}
    \State $supportingNodes \gets []$
    \ForAll{$node \in \text{AllNodes}$}
        \If{$node$ has voted for blocks stemmed from $B_{k'}$ at height $k$}
            \State Append $node$ to $supportingNodes$
        \EndIf
    \EndFor
    \State \textbf{return} $supportingNodes$
\EndFunction

\Function{VoteCount}{$B_{k'}$, $\operatorname{VV}_j^{(k)}$}
    \State \textbf{return} Count of votes for $B_{k'}$ and blocks stemmed from $B_{k'}$ in $\operatorname{VV}_j^{(k)}$
\EndFunction

\Function{MissingVotesCount}{$k''$, $G_j^{(k)}$}
    \State \textbf{return} Count the number of votes from up until (including) height $k''-1$ not shown in $G_j^{(k)}$
\EndFunction
\end{algorithmic}

\end{breakablealgorithm}

\noindent \textbf{Rationale behind Eq.~\ref{acc7}}: Assume that a block $B_{k'}$ of consensus height $k'$ is consensually accepted at height $k>k'$ (Eq.~\ref{acc7} is satisfied). When $\operatorname{VC}_{j}^{(k)}[B_{k'}]> \operatorname{VC}_{j}^{(k)}[B'_{k'}(j)]+\operatorname{VC}_{i}^{(k)}[U_{k}]-\operatorname{VC}_{i}^{(k)}[U_{k'}]$, any node $j$ in $NG$ would have voted for the branch of $B_{k'}$ at height $k$. Failure to do so would imply adversarial characteristic (C2) and is therefore not in $NG$.
Despite this, network delays might cause some node $i$ in $NG$ to send votes for height $k+1$ before receiving some other votes at $k$ from nodes in $NG$. This may potentially affect node $i$'s vote for height $k+1$ because in $G_i^{(k+1)}$ there could be less than $N-f$ nodes voted for blocks stemming from $B_{k'}$ at $k$. To ensure continued support for $B_{k'}$ at $k+1$, it must be demonstrated that for any vote of height $k+1$ linked to any $N-f$ votes at vote height $k$, $B_{k'}$ remains the root of the most supported branch of $k'$ in its vote count and this support is greater than the second most supported plus the missing votes.

In the worst-case scenario, only $f+1$ votes of height $k$ from nodes in $NG$ are linked by a new vote of node $i$ of height $k+1$. This is because $N=3f+1$ and $NG$ contains $2f+1$ nodes. As mentioned, nodes in $NG$ must have voted in $B_{k'}$'s branch at height $k$. After receiving $N-f$ votes at height $k$, if $\operatorname{VC}_{i}^{(k)}[B_{k'}] +f+1 > \operatorname{VC}_{i}^{(k)}[U_{k}]-\operatorname{VC}_{i}^{(k)}[U_{k'}] + \operatorname{VC}_{i}^{(k)}[B'_{k'}(i)] +f$, node $i$ will still vote in $B_{k'}$'s branch at $k+1$. This is because, if node $i$ discovers all missing votes and finds that they support $B'_{k'}(i)$, and there are $f$ votes supporting $B'_{k'}(i)$ in $G_{i}^{(k+1)}$, the branch rooted by block $B_{k'}$ remains the most supported.

If adversarial characteristic (C1) is detected in the future, double votes are not counted. Assuming the votes were counted in $\operatorname{VC}_{i}^{(k)}[B_{k'}]$ already, then it should satisfy $\operatorname{VC}_{i}^{(k)}[B_{k'}] +f+1 -t > \operatorname{VC}_{i}^{(k)}[U_{k}] -\operatorname{VC}_{i}^{(k)}[U_{k'}] + \operatorname{VC}_{i}^{(k)}[B'_{k'}(i)] +f$, where $t=(k-k')\times f$.

Thus, once Eq.~\ref{acc7} is satisfied, nodes cannot vote for blocks in other branches, establishing the uniqueness of acceptance for block $B_{k'}$. Lem.~\ref{k2-new} is a formal statement of the safety of this algorithm. 

\begin{theoremEnd}[end, restate]{Lemma}
\textbf{(Safety)} A block $B_{k'}$ accepted in consensus satisfying Eq.~(\ref{acc7}) is unique, and there cannot exist another block accepted in consensus in a branch different from the one stemmed from $B_{k'}$ at any time.
\label{k2-new}
\end{theoremEnd}
\begin{proofEnd}
    \textbf{Assumption:} Assume that our protocol correctly implements and enforces the rules stated in the consensus mechanism, especially the handling of node votes, double-voting detection, and compliance with voting rules according to vote counts.

\textit{Step 1: Assume for Contradiction}

Assume, for the sake of contradiction, that there exists another block \( B''_{k'} \) different from \( B_{k'} \), which is also accepted in consensus at the same height \( k' \), despite \( B_{k'} \) having been accepted following the consensus criteria.

\textit{Step 2: Analysis of the Consensus Criteria}

Under the consensus rules:
\begin{enumerate}
\item All nodes in a subset \( NG \) of size at least \( N-f \) (where \( f \) is the number of possible faulty nodes) have voted for \( B_{k'} \) such that:
   \[ \operatorname{VC}_{j}^{(k)}[B_{k'}] - t > \operatorname{VC}_{j}^{(k)}[B'_{k'}(j)] + \operatorname{VC}_{j}^{(k)}[U_{k}] - \operatorname{VC}_{j}^{(k)}[U_{k'}] - 1 \]
   where \( t \) accounts for potential double-voting between \( k' \) and \( k-1 \).
\item The condition essentially states that the support for \( B_{k'} \) exceeds any potential support for any other block, taking into account uncounted votes and potential double votes.
\end{enumerate}

\textit{Step 3: Security Reduction}
\begin{itemize}
\item By the consensus condition, for \( B_{k'} \) to be accepted, it must be the most supported block at height \( k' \) after considering all compliant votes.
\item If \( B''_{k'} \) were also accepted under these conditions, it would imply that \( B''_{k'} \) too met the threshold in Eq. (\ref{acc7}), contradicting our initial validation that \( B_{k'} \) had decisively more support.
\item Such a scenario would suggest that the system has either failed to detect and discard invalid or non-compliant votes (such as those involved in double-voting) or has incorrectly calculated the vote totals. This would violate the integrity of the voting mechanism designed to prevent exactly such an occurrence.
\end{itemize}

\textit{Step 4: Implication of the Assumption Being False}

\begin{itemize}
\item If the lemma were false, it would imply that the system’s core security features (accurate vote counting, compliant vote validation, and correct execution of consensus protocol) are flawed.
\item Since the protocol relies on these features to ensure network security and integrity, which are assumed to be reliable, the assumption that two different blocks can be accepted at the same height must be false.
\end{itemize}

\textbf{Conclusion:} The acceptance of two different blocks \( B_{k'} \) and \( B''_{k'} \) in the same consensus height would directly contradict the enforcement of the voting and consensus rules. Therefore, by security reduction, if our security mechanisms are functioning as intended, no such scenario can occur. Hence, the lemma holds under the conditions set by our protocol, ensuring the uniqueness of the accepted block at any consensus height.
\end{proofEnd}

For better understanding, Appx.~\ref{G} shows an online visualization demonstration of our protocol.

\subsection{Liveness of Consensus Derivation}
To guarantee consensus-derivation liveness, we must exhibit that the number of blocks ($m$) at a specific consensus height supported by nodes---by casting votes for blocks stemming from them---monotonically decreases until all the nodes have voted for the same branch. The asynchronous network stipulates that the honest message delay is arbitrary but finite. It does not address to what extent Byzantine nodes can manipulate these delays. Here, we detail different situations.

\subsubsection{Practical liveness}\label{4.6.1}
It is generally known that asynchronous BFT consensus protocols require unpredictable randomness 
so that the adversary can only adapt to the randomness when it is too late to influence the outcome.

Random message delays introduce entropy into the system. In our design, each node derives its voting decision from the first $N-f$ compliant votes of the previous height it has received, akin to utilizing a weak common coin~\cite{Abraham2022Asynchronous}, assuming that the adversary has no control over which $N-f$ votes each node will receive first. This consequently reduces $m$ when some branches no longer remain the most supported in any vote's vote count. Ultimately, only one branch remains. This approach possesses several key properties:
\begin{itemize}[leftmargin=0.2cm, itemindent=-0cm,itemsep=2pt,topsep=0pt,parsep=0pt]
\item \textbf{Randomness:} If the first $N-f$ compliant votes of the last vote height received by an honest node are random, it can be seen as equivalent to this honest node votes 
for one among $m$ blocks 
at random. 
\item \textbf{$\epsilon$-correct:} For any $b \in [1, m]$, with probability $\epsilon$, all nodes output $b$, where  $\epsilon \geq\frac{\frac{N-f}{m}}{N}$. This probability arises because, irrespective of the votes from previous heights, a block among the $m$ blocks must gain at least $\frac{N-f}{m}$ votes in order to potentially be the most supported in a set of $N-f$ votes.
\item \textbf{Monotonicity:} Assume a strong adaptive adversary who sees the internal state of all nodes and the content of all messages sent, and it makes its own vote decisions accordingly. It could influence the coin value  (the most supported block according to one's vote count). But, the adversary cannot decide the coin value without considering $f+1$ previous honest votes, which monotonically reduces the number of options for the adversary round by round. It, therefore, monotonically converges to one branch eventually.
\end{itemize}



\begin{theoremEnd}[end,restate]{Lemma}
 \label{NS} 
\textbf{(Normal case)} Consider $N$ nodes voting for one of $m \leq N$ distinct input values. In each round, a node can observe $N-f$ votes from the previous round, and each vote is recursively linked to the $N-f$ votes from the previous round. The process of convergence can be modeled as a discrete-time Markov chain (DTMC) with $m$ states, representing the number of remaining input values. In this model: \textbf{Worst Case}: The convergence time for all nodes to vote for the same input value is $O(m)$ rounds. \textbf{Amortized Case}: In the average case, the system converges in $O(\log m)$ rounds.
\end{theoremEnd}

\begin{proofEnd}
We model the process of convergence as a discrete-time Markov chain (DTMC) with state space $\{1, 2, \dots, m\}$, where each state $x_r = x$ represents the number of distinct input values remaining at round $r$. The absorbing state is $x = 1$, where consensus is reached, and no further transitions occur.

\textbf{State Space and Transitions:}
\begin{itemize}
    \item The process begins at state $x_r = m$, where all $m$ distinct input values are still in play.
    \item The system transitions between states based on whether one or more input values are eliminated in each round. A transition from $x_r = x$ to $x_{r+1} = x-j$ represents the elimination of $j$ input values in that round, where $j \in \{1, 2, \dots, x-1\}$.
    \item The absorbing state is $x = 1$, where all nodes vote for the same input value and convergence is reached.
\end{itemize}

\textbf{Transition Probabilities:}
Let $P(x \to x)$ be the probability that no progress is made in a round, meaning no input value is eliminated. Let $P(x \to x-j)$ be the probability that $j$ input values are eliminated in a round.

\begin{enumerate}
    \item \textbf{No Progress Probability $P(x \to x)$}: 
    This occurs when the nodes \emph{nearly equal supporting the remaining $x$ input values}, and no clear majority emerges. The probability of no progress is higher when the votes are more evenly distributed. Based on binomial distribution approximations, \emph{a very lost upper bound} probability can be written as:
    \[
    P(x \to x) \approx \frac{1}{\sqrt{2 \pi \frac{N-f}{4}}}
    \] The rational behind this probability is that consider the case where there are two equally supported branches rooted by block $A$ and $B$, and each has $\frac{N}{2}$ votes. Each node samples $N-f$ votes from this set. Afterwards, the probability that exactly $\frac{N}{2}$ nodes vote for each branch (thus no change in the poll) can be approximated using the \textit{binomial distribution}: \[p(o) = \binom{N-f}{o} \left( \frac{1}{2} \right)^o \left( \frac{1}{2} \right)^{(N-f-o)} = \binom{N-f}{o} \left( \frac{1}{2} \right)^{N-f}\]where $o = \frac{N-f}{2}$. The probability (very loose upper bound) is thus approximately:$p_{(\text{no progress})} \approx \frac{1}{\sqrt{2 \pi \frac{N-f}{4}}}$.

    \item \textbf{Progress Probability $P(x \to x-j)$}: 
    The probability that $j$ input values are eliminated in the round depends on how biased the vote distribution is. As $x$ decreases (fewer input values remain), the likelihood of eliminating multiple input values at once increases if the vote distribution skews towards one or more dominant input values.
    
      The probabilities for eliminating $j$ input values in a single round can be modeled as a decreasing geometric-like distribution where the probability of eliminating more input values declines as $j$ increases:
    \[
    P(x \to x-j) = \alpha_j \cdot P(\text{progress}), \quad \sum_{j=1}^{x-1} \alpha_j = 1
    \]
    where $\alpha_j$ is a weight assigned to the probability of eliminating $j$ input values. For simplicity, these weights can reflect the natural tendency that eliminating fewer input values is more likely than eliminating more in each round.
\end{enumerate}

\textbf{Transition Matrix:}
The transition matrix $P$ for this Markov chain is defined as follows, where $P_{ij}$ represents the probability of transitioning from state $i$ to state $j$:
\[
\small
P = \begin{pmatrix}
1 & 0 & 0 & \dots & 0 & 0 \\
P(2 \to 1) & P(2 \to 2) & 0 & \dots & 0 & 0 \\
P(3 \to 1) & P(3 \to 2) & P(3 \to 3) & \dots & 0 & 0 \\
\vdots & \vdots & \vdots & \ddots & \vdots & \vdots \\
P(m \to 1) & P(m \to 2) & P(m \to 3) & \dots & P(m \to m-1) & P(m \to m) \\
\end{pmatrix}
\]

\textbf{Expected Time to Absorption:}
The goal is to calculate the expected time to reach the absorbing state $x = 1$, starting from state $x = m$. Let $t_x$ represent the expected number of rounds to reach state $x = 1$ starting from state $x$. The recurrence relation for $t_x$ 
is:

\[
t_x = \frac{P(x \to x) + \sum_{j=1}^{x-1} P(x \to x-j) + \sum_{j=1}^{x-1} P(x \to x-j) \cdot t_{x-j}}{1 - P(x \to x)}
\]

This equation indicates that the expected time $t_x$ depends on either staying in the same state (with probability $P(x \to x)$) or transitioning to lower states, reducing multiple input values at once. The summation term accounts for all possible transitions, where $j$ input values are eliminated in a round.

Rearranging the terms:

\[
t_x = \frac{1 + \sum_{j=1}^{x-1} P(x \to x-j) \cdot t_{x-j}}{P(\text{progress})}
\]

By solving this recurrence relation starting from $t_1 = 0$ (the absorbing state), we can calculate the expected time to convergence. 

\textbf{Worst-Case Convergence Time:} 
In the worst case, when the votes are nearly uniformly distributed, $P(x \to x)$ remains high, and progress is slow. The recurrence relation shows that when $P(x \to x)$ is significant, the expected time $t_x$ increases linearly with $x$, leading to an overall convergence time of $O(m)$. 

\textbf{Amortized Case Convergence Time:} 
In the average case, as $x$ decreases, the vote distribution becomes less uniform, and the probability of eliminating multiple input values in a round increases. Once a majority input value emerges, recursive vote linking amplifies its support, leading to exponential convergence. The expected time to consensus in this case is $O(\log m)$.

\end{proofEnd}

Appx.~\ref{aLCR} shows a simple simulation that faithfully represents the voting rounds before convergence. With $N=100$ and $m=100$, convergence happens within $4$ voting rounds.


\subsubsection{Cope with adversary controlling network delays}\label{4.6.2}

\label{aced}\label{sec:adversarial_control_vote_delivery}\label{am}

The randomness of our protocol may be compromised if an adversary can control which set of $N-f$ votes are first received by each and every node.

\begin{theoremEnd}[end, restate]{Lemma}
\textbf{(Loss of Liveness in Adversarial Settings)} The liveness of the consensus-derivation process is compromised when an adversary controls network delays entirely.
\label{liveness}
\end{theoremEnd}

\begin{proofEnd}
Assume that all $m$ blocks receive $\frac{N}{m}$ votes at the first voting height. Ideally, each node recognizes multiple blocks as equally supported; subsequently, each node votes randomly. The least-supported blocks will be discarded, reducing $m$. However, nodes only have access to partial voting history, which could result in divergent perspectives. For example, in an $N=12, f=3$ system with $m=4$ blocks ($A, B, C, D$), when the global view of votes is $AAABBBCCCDDD$, nodes might hold different local views such as $AAABBCCDD$ and $BBBAACCDD$, depending on the first $N-f$ compliant votes they receive. 
If the adversary controls the set of votes received by the honest nodes, it can manipulate them to consistently vote for the same branch by carefully selecting which $N-f$ votes they receive first. This attack can prevent the reduction of $m$, thus hindering liveness. 
\end{proofEnd}

We adjust our protocol's randomness during attacks.

\noindent\textbf{Indications of potential loss of liveness.} A potential hindrance to liveness only happens when at least $N-f$ votes from height $k-1$, cast by any node $j$, \emph{uniformly} satisfy the condition $\operatorname{VC}_{j}^{(k-1)}[B_{k'}(j)] > \operatorname{VC}_{j}^{(k-1)}[B'_{k'}(j)]$. But, it must also be true that $\operatorname{VC}_{j}^{(k-1)}[B_{k'}(j)] < \operatorname{VC}_{j}^{(k-1)}[B'_{k'}(j)] + \operatorname{VC}_{j}^{(k-1)}[U_k] - \operatorname{VC}_{j}^{(k-1)}[U_{k'}]$, where node $j$ voted for a block at height $k-1$ stemming from $B_{k'}(j)$. $B'_{k'}(j)$ is the block at height $k'$ that, according to $\operatorname{VC}_j^{(k-1)}$, has the most vote count apart from $B_{k'}(j)$. 

\noindent\textbf{Regain liveness.} 
Honest node $i$ execute the following steps if all of these conditions are met: (1) node $i$ receives $N-f$ votes of height $k-1$, (2) each vote $V_{j}^{(k-1)}$ was cast for a block stemming from $B_{k'}(j)$, and $\operatorname{VC}_{j}^{(k-1)}[B_{k'}(j)] > \operatorname{VC}_{j}^{(k-1)}[B'_{k'}(j)]$, but $\operatorname{VC}_{j}^{(k-1)}[B_{k'}(j)] < \operatorname{VC}_{j}^{(k-1)}[B'_{k'}(j)] + \operatorname{VC}_{j}^{(k-1)}[U_k] - \operatorname{VC}_{j}^{(k-1)}[U_{k'}]$, and (3) $k-k' > \log_2(N)$.
    \begin{enumerate}[leftmargin=0.4cm, itemindent=-0cm,itemsep=2pt,topsep=0pt,parsep=0pt]
    \item Flip the binary strong common coin~\cite{mostefaoui2014signature} at height $k$.
    \item If the result is 0, it votes for a block stemming from $B_{k'}(i)$, otherwise, it votes for a block stemming from $B'_{k'}(i)$. $B_{k'}(i)$ and $B'_{k'}(i)$ are the first and second most supported blocks according to $\operatorname{VC}_i^{(k)}$. $V_{i}^{(k)}$ must attach the verification of the coin flip to validate its vote and avoid being classified as a non-compliant vote.
    \end{enumerate}
\noindent \textit{Remarks:} Only \emph{one} coin flip is needed at height $k$, rather than one coin flip per node $i$. Each node broadcasts a threshold signature, with $f+1$ as the threshold. Thus, a node generates only one message per coin flip. A coin flip incurs $O(N)$  messages across the system. The amendment is only activated when $k-k' > \log_2 (N)$, where $\log_2(N)$ is the expected rounds until convergence with random message delays, assuming all $N$ nodes proposed a block at height $k'$ (see Lem.~\ref{NS}). It is trivial to prove that \( m \) halves after each round in amortization, thus necessitating $log_2(m)$ voting rounds. 

\begin{theoremEnd}{Lemma} The amortized and worst voting rounds until convergence in adversarial settings are $log_2(mN)$ and $N+log_2(m)$ respectively.
\end{theoremEnd}

\begin{theoremEnd}[end, restate]{Lemma}
The safety of the liveness regain method and the liveness of the consensus-derivation process are both guaranteed.
\label{ame}
\end{theoremEnd}

\begin{proofEnd}
\noindent \textbf{Safety of the established consensus:} Though a node may oscillate between different blocks, this inconsistency will not affect the already accepted blocks because if $B_{k'}$ is already accepted before or at $k$, $\operatorname{VC}_{i}^{(k)}[B_{k'}] \leq \operatorname{VC}_{i}^{(k)}[B'_{k'}(i)] + \operatorname{VC}_{i}^{(k)}[U_{k}] - \operatorname{VC}_{i}^{(k)}[U_{k'}]$ will not occur.

\noindent \textbf{Safety of adversary controlling the delays:} An adversary may control which $N-f$ votes are received by each honest node and can trigger indication of potential loss of liveness. However, the strong common coin is flipped later. Thus, the adversary cannot influence the random vote decisions.

\noindent \textbf{Liveness:} The randomness of voting that assists in breaking away from the loss of liveness can be provided by a local coin. However, allowing not voting for the most supported branch enables a strong adaptive adversary to escape from the irreversible convergence enforced by the vote graph if it can observe the votes of others before casting its own. By implementing a strong common coin, the adversary must follow the unpredictable coin value. This provides an unbiased randomness necessary to ensure liveness.

If fewer than $N-f$ nodes participate in the coin flip, the coin flip will not be generated. There must exist some honest node $i$, where the most supported branch $B$ satisfies the condition $\operatorname{VC}_i^{(k)}[B] > \operatorname{VC}_i^{(k)}[B'] + \operatorname{VC}_i^{(k)}[U_k] - \operatorname{VC}_i^{(k)}[U_{k'}]$ for any other branch $B'$. In this case, node $i$ will vote for branch $B$ at height $k$ instead of participating in the coin flip. Then, the other honest nodes will eventually receive $V_i^{(k)}$ according to the assumption that the adversary cannot delay messages indefinitely. They can then link their votes to a different set of $N-f$ votes from the previous height and vote for the most supported branch to their knowledge after acquire the missing votes from $G_i^{(k)}$. 
\end{proofEnd}

\begin{theoremEnd}[end, restate]{thm}
\textbf{(Assurance of Liveness)}
The protocol guarantees liveness even if the adversary control network delays.
\label{thm:liveness}
\end{theoremEnd}
\begin{proofEnd}
There are two cases, which we analyze separately below:

Case A (practical liveness, the adversary does not control the delays of at least $N-f$ nodes): Because each node receives a random set of $N-f$ votes, it can be seen as flipping a weak random coin, as discussed in Sec.~\ref{4.6.1}. This will monotonically reduce the input values until only one remains.

Case B (The adversary constantly controls the delays of at least $N-f$ nodes): If the condition $\operatorname{VC}_j^{(k-1)}[B_{k'}(j)] < \operatorname{VC}_j^{(k-1)}[B'_{k'}(j)] + \operatorname{VC}_j^{(k-1)}[U_k] - \operatorname{VC}_j^{(k-1)}[U_{k'}]$ is uniformly satisfied by at least $N-f$ votes from height $k-1$ cast by any node $j$, and $k-k'>log_2 (N)$, indicating potential for multiple supported branches (as detailed in Sec.~\ref{4.6.2}), honest nodes initiate a coin flip. The coin flip is a strong common coin mechanism that produces a common random value that cannot be influenced by the adversary once initiated. The outcome of the coin flip dictates which branch the honest nodes will vote for at height $k$, providing a way to break the tie uniformly across all honest nodes. Since the adversary cannot predict or manipulate the outcome of the coin flip once it is triggered, the coin flip ensures that, after each round, the chance of all honest nodes aligning on the same branch increases as it can be seen as the nodes receiving a random set of $N-f$ votes of the last height, akin to Case A.

Therefore, the eventual convergence is guaranteed.
\end{proofEnd}
\begin{theoremEnd}[end,restate]{thm}
\textbf{ (Agreement and Termination)}
At some point in the consensus process, a single block will be irreversibly accepted, provided that unbiased randomness is guaranteed.
\label{k7}
\end{theoremEnd}
\begin{proofEnd} 
We first assume all nodes adhere to the voting rules defined in Sec.~\ref{lba}. Thm.~\ref{thm:liveness} shows that they will converge to the same branch despite the presence of strong adaptive adversary. After this process, Eq.~\ref{acc7} will be achieved. The convergence and exclusivity of the votes guarantee irreversible block acceptance, as suggested in Lem.~\ref{k2-new}.

Now, consider what happens if violating the voting rules:

\begin{itemize}
    \item \textbf{Non-compliant votes:}
    Lem.~\ref{k1} proves that any non-compliant vote, i.e., those exhibiting characteristic (C2) of adversarial behavior, is disregarded by honest nodes upon receiving them. 

    \item \textbf{Double voting:}
    Concerning double voting (adversarial characteristic (C1) and a breach of honest model~\ref{p1}), Lem.~\ref{k2-new} proves that the criteria for block acceptance already consider and mitigate the effects of double voting. Eq.~\ref{acc7} suggests the system's ability to achieve consensus despite double voting.
\end{itemize}

Hence, non-compliant votes and double voting do not influence the outcome. If the liveness of consensus-derivation is secured, agreement and termination are assured.
\end{proofEnd}

\subsection{Optimizations}
\subsubsection{Speeding up consensus derivation process}

Consider a scenario where there are only two blocks, \(B_{k'}\) and \(B'_{k'}\), at consensus height \(k'\), and the adversary remains silent (i.e., does not participate in voting). Eq. \ref{acc7} can be transformed as follows:
\begin{equation}
\begin{aligned}
\forall j \in NG: \quad 
(k - k') \times (2f+1) - \operatorname{VC}_{j}^{(k)}[B'_{k'}]-(k - k') f
> \\ \operatorname{VC}_{j}^{(k)}[B'_{k'}] + (k - k')f - 1
\label{xxx}
\end{aligned}
\end{equation} 

Eq.~\ref{xxx} can be simplified to:

\begin{equation}
\begin{aligned}
\forall j \in NG: \quad \quad \quad 
(k - k') > 2 \operatorname{VC}_{j}^{(k)}[B'_{k'}] - 1
\label{xxx1}
\end{aligned}
\end{equation}

Eq. \ref{xxx1} indicates that at least \(2 \operatorname{VC}_{j}^{(k)}[B'_{k'}]\) rounds of voting are required to reach a consensus, even though the nodes may have already converged to the branch of \(B_{k'}\) in the duration. This round number is not ideal.

\noindent\textbf{Acceleration:} To accelerate the process, we modify Eq. \ref{acc7} to:

\begin{equation}
\begin{aligned}
\forall j \in NG: \quad \quad \quad 
\operatorname{VC}_{j}^{(k)}[B_{k'}]\{k-1\} - f  > \\ \operatorname{VC}_{j}^{(k)}[B'_{k'}(j)]\{k-1\} + \operatorname{VC}_{j}^{(k)}[U_{k}]\{k-1\} - 1
\label{acc8}
\end{aligned}
\end{equation}

Here, \(\operatorname{VC}_{j}^{(k)}[B_{k'}]\{k-1\}\) denotes the sum of votes specifically for blocks at consensus height \(k-1\) in the branch of \(B_{k'}\). \(\operatorname{VC}_{j}^{(k)}[U_{k}]\{k-1\}\) represents the missing votes at vote height \(k-1\). Eq. \ref{acc8}, which focuses solely on votes at height \(k-1\), remains valid and can be seen as a projection of Eq. \ref{acc7} because:

1) Similar to Eq. \ref{acc7}, Eq. \ref{acc8} holds only if the branch rooted by \(B_{k'}\) is the most supported branch; otherwise, it implies that the votes are non-compliant, which is impossible as they have been counted to $\operatorname{VC}_{j}^{(k)}$.

2) The parameter \(t\) can be substituted with \(f\) because, once a double-voting vote is detected, all subsequent votes in the vote chain of the voter are disregarded; at most \(f\) nodes may have double-voted, resulting in at most \(f\) votes at height \(k-1\) that are not being counted. 

In the same scenario with silent Byzantine nodes and only two blocks \(B_{k'}\) and \(B'_{k'}\) at consensus height \(k'\), Eq. \ref{acc8} can be simplified to \(\operatorname{VC}_{j}^{(k)}[B'_{k'}]\{k-1\} < 1\). It implies that,  consensus is reached as fast as honest nodes have converged to the branch of \(B_{k'}\) before $k$ and  \(\operatorname{VC}_{j}^{(k)}[B'_{k'}]\{k-1\} = 0\).

Thus, replacing Eq.\ref{acc7} with Eq.\ref{acc8} will accelerate the consensus derivation process in light of silent adversary.

\subsubsection{Smooth vote height advancement}\label{relaxing}
In the design described so far, a node will vote for a block of the next height when it receives $N-f$ compliant votes from the current height. This design has the following two problems: (1) The slow compliant votes may never be linked by the fast compliant votes. (2) The gap between the consensus height and the vote height can be significant. This is because the consensus height advances smoothly with a new leader proposing a block at the beginning of each $\Delta BI$. The timing of the vote height progression depends on the slowest vote among the $N-f$ compliant votes of the current height. As message delays are uncontrollable, the speed of vote height progression can be much slower than that of consensus height progression.

To address these, we introduce a relaxed design, enabling nodes to vote periodically, regardless of whether they have received $N-f$ votes. 
We define:

1). Each vote indicates a checkpoint height ($cph$). For votes in the first vote height, $cph=0$.

2). $cph$ indicated in $V_j^{(k)}$ is defined as the latest vote height where at least $N-f$ votes of this vote height sharing the same $cph$. $cph$ for $V_j^{(k)}$ can be determined using $G_j^{(k)}$.

\begin{figure}[h!]
    \centering\vspace{-0.1cm}
    \includegraphics[width=0.41\textwidth]{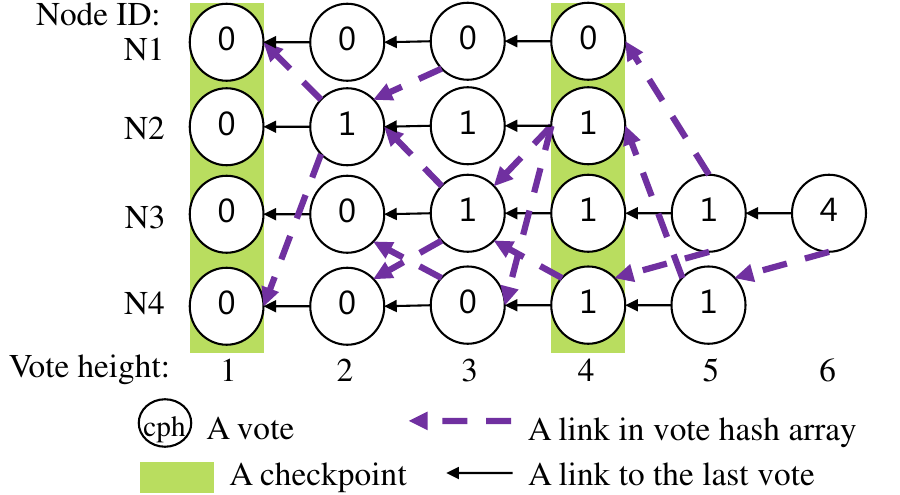}\vspace{-0.2cm}
    \caption{The example showing the process of updating $cph$}
    \label{fig:relaxed-example}\vspace{-0.4cm}
\end{figure}

We illustrate this with an example in Fig.~\ref{fig:relaxed-example}. This system has four nodes, all initially sharing $cph=0$. Node $N2$ knows three votes at vote height 1, prompting it to update its $cph$ to $1$. Subsequently, Node $N3$ receives the vote from $N2$ of vote height 2 and realizes that all four nodes with $cph=0$ have voted at height 1. As a result, $N3$ updates its $cph$ to 1 at vote height 3. Similarly, $N4$ updates its $cph$ to 1 at vote height 4. At vote height 6, Node $N3$ observes that at least $N-f$ nodes share $cph=1$ at vote height 4, prompting it to update its $cph$ to 4.

We enact amendments to the voting rules as follows:

1). Except for votes at vote height 0, a vote must link to the preceding vote of the voter (still maintaining its own vote chain). It may include links to any number of votes from different voters of the previous vote heights (not necessarily of the last vote height). It should not link two votes from one voter at one time.

2). Nodes are permitted to switch branches only when a larger $cph$ is determined based on their vote graphs; otherwise, their votes are non-compliant votes.

3). Eq.~\ref{acc1} (for detecting non-compliant votes) is only used for evaluating the compliance of the vote when a greater $cph$ is determined in this vote.

4). $cph$ is used as $k$ for Eq.~\ref{acc8}. Only when $cph$ is updated do they check if a block has been consensually accepted.

Eq.~\ref{acc8} remains valid in this relaxed design as nodes cannot switch to other branches until they have updated their $cph$. In other words, all nodes voted in the branch rooted by blocks $B_{k'}$ or $B'_{k'}$ will continue to vote in the same branch until $cph$ is updated. Thus, any vote count gap between $B_{k'}$ and $B'_{k'}$ will only extend before $cph$ is updated. Eq.~\ref{acc8} will not be reversed. If adversarial votes do not link to the honest votes, they cannot reach a new $cph$ and they are being locked in their original choices (similar to being disregarded).  

It is trivial to prove that the relaxed design, which allows nodes to move on to new vote heights earlier, will accelerate block acceptance when message delays are random (i.e., the slowest honest node is not always the same).

\section{Analysis}
\subsection{Simplified Complexity}
In our protocol, nodes do not broadcast an acknowledgment to all when receiving a block. If a node misses a block, the block can be discovered via the vote graph. Therefore, a view change requires just the leader to simply broadcast a block with $O(N)$ message complexity. Each node votes once per height; since there are only $N$ votes, the  message count remains at $O(N)$. Furthermore, since each broadcast incurs $O(N)$ communication, the total complexity of vote broadcasting per instance is $O(\theta N^2)$, where $\theta$ is the number of voting rounds until consensus is reached. Our protocol thus aligns with partially synchronous consensus protocols at an $O(N^2)$ complexity level. Optimizing broadcasting through methods such as the gossip protocol can reduce it to \(O(\theta N \log N)\), or a leader-collecting system, like in HotStuff, can reduce our protocol's complexity to \(O(\theta N)\). In a leader-collecting system, leaders collect the votes from $N-f$ nodes and broadcast them as one message to all nodes (HotStuff uses threshold signatures to reduce data cost). In our context, leaders may control which $N-f$ votes to broadcast, potentially influencing subsequent voting. But, liveness for consensus-derivation is ensured with the amendment in Sec.~\ref{sec:adversarial_control_vote_delivery}, as verified in Lem.~\ref{ame}.
It is trivial to prove that the amendment in Sec.~\ref{sec:adversarial_control_vote_delivery} can use a leader-collecting broadcasting as well.

\noindent \textbf{Comparison with existing BFT asynchronous consensus protocols for SMR.} 
Existing protocols often involve $O(N^3)$ message complexity and at least $O(N^2)$  messages. Our protocol may reduce both to $O(N)$ as discussed.

\noindent \textbf{Comparison with partially synchronous leader-based protocols.}
Traditional BFT protocols, like PBFT~\cite{10.5555/296806.296824}, involve $O(N^2)$ complexity for pre-vote alignment, with each node re-broadcasting received input values. In contrast, HotStuff assigns sole broadcasting responsibility to the leader, who then seeks a collective threshold signature from the nodes, simplifying the process to $O(N)$. Our protocol, requiring no pre-vote alignment, allows a leader to directly broadcast blocks, achieving $O(N)$ broadcasting complexity without threshold signatures. In short, our protocol runs in asynchronous settings but matches the message complexity of HotStuff, which runs in partially synchronous settings. 



\noindent \textbf{Comparison with classical strong consensus protocols.}
Classical method (Protocol 5 in \cite{10.1145/872035.872066}) entails three broadcasting rounds per voting round---initial vote broadcasting, subsequent two rounds confirming whether input values reach sufficient votes---our protocol simplifies to a single broadcast per round, except under specific conditions outlined in Sec.~\ref{am}. Moreover, by allowing nested instances and concurrent voting rounds, our protocol supports smooth blockchain progress despite slow honest nodes.


\subsection{Reduced Data Cost}
The size of a vote in our system is determined by the number of votes from the previous heights this vote linked to. In the most data-costly case, a vote may link to $N$ previous votes. Assuming that the hash function is SHA-256, this vote occupies more than $256 \times N$ bits. If in a PKI-enabled network, each message is associated with additional $512$ bits (e.g., secp256r1) for a digital signature. To synchronize all votes at a given height, a node needs storage exceeding $(32 \times N+64) \times N$ bytes, presenting a substantial storage cost. When $N=1000$, it results in a significant $30.57 MB$ of data. In response to this challenge, we propose an optimization where each vote includes an $N$-bit string representing votes received from other nodes. If the $i$-th bit is 1, then the voter has received the vote from node $i$. A bloom filter is attached to encapsulate these votes. When a node receives a vote, it goes through the $N$-bit string and builds the links from previously received votes if the votes are inside the bloom filter. When a vote cannot be located through the bloom filter, it can be acquired from the voter. Honest voters respond accurately. If an adversarial voter fails to respond, the vote is safely disregarded.

\subsubsection{Cost analysis for using bloom filter}\label{datacost}
\begin{figure}[h!]
    \centering
    \includegraphics[width=0.46\textwidth]{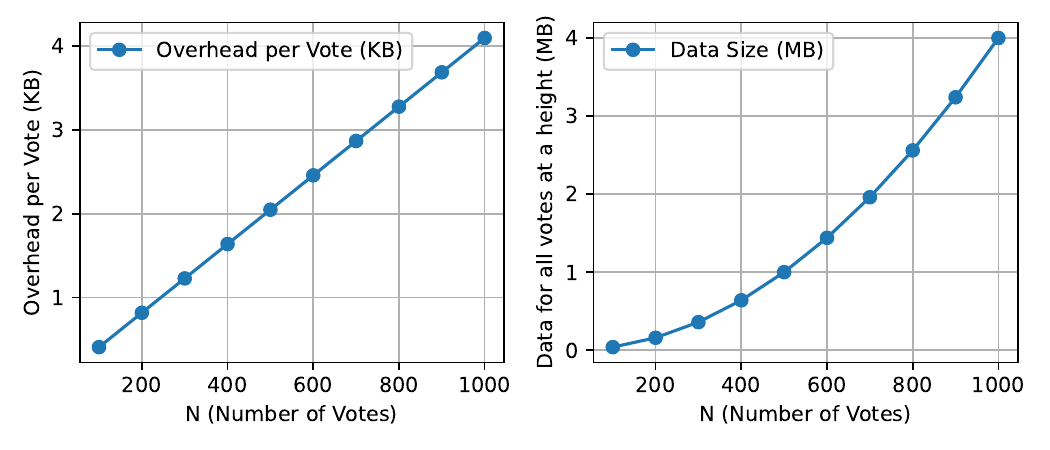}\vspace{-0.4cm}
    \caption{The vote size and the data required for storing all votes at a block height with $p=10^{-7}$ and different $N$}\vspace{-0.5cm}
    \label{fig:vote_size}
\end{figure}

Considering a false positive rate of $p=10^{-7}$, a level of precision is suitable for critical applications. Consider $N=1000$, using the formula $a = -\frac{N \ln(p)}{(\ln(2))^2}$, $ a \approx 33,568.55$. 
Therefore, the recommended bloom filter size ($a$) for $N = 1000 $ and $ p = 10^{-7} $ is approximately only $33569$ bits.  
In this case, the bloom filter introduces an overhead of approximately $4.1$ KB per vote, significantly mitigating the excessive data costs associated with the initial design. The data for storing \emph{all} votes at a given height is more than $4.162$ MB after considering the vote signatures.  Fig.~\ref{fig:vote_size} shows the data size with $p=10^{-7}$ and different $N$, suggesting the data required for synchronizing the votes in our protocol is tolerable. 

\subsubsection{Comparison with other asynchronous blockchains}\label{sig}
Our protocol's data cost is minimal compared to both ACS-based and DAG-based protocols, despite the integration of hash values in votes (stored using bloom filter.) To compare with our protocol, we analyze the following two factors:

\noindent \textbf{Overlapped transactions:}  In both ACS-based and DAG-based protocols, at least honest nodes will reliably broadcast their set of transactions to the network. Overlapping among the transaction sets is unavoidable as users would not trust any single node and will submit their transactions to more nodes to prevent censorship. So, the data redundancy of overlapping transactions will likely incur significant data costs. But, even under an unlikely scenario where each node in our protocol proposes a block at every height, our data cost is less. A node of our protocol only downloads \emph{one} block per height and then vote for it, it downloads another block only when switching branches or having downloaded an invalid block.

\noindent \textbf{Limited network topologies:} Existing protocols incur an \(O(N^2)\) communication cost for downloading  messages per node per instance. This cost occurs in fully-meshed authenticated point-to-point channels. It is worth noting that some ACS-based protocols employ threshold signatures in certain phases of their execution; but most messages are not signed. As a result, they cannot be multicast via gossip protocols and must be sent directly from the source to the destination. The eventual message delivery often involves repeating data transmissions until an acknowledgment is received. If Dumbo2 signs messages, each node will download over \(64 \times N^2\) extra bytes for message signatures. With \(N = 1000\), this is over 61 MB. A similar pattern is exhibited in DAG-based protocols. The vote cost discussed in Sec.~\ref{datacost} already accounts for signatures, showing our protocol practically supports network topologies that use message forwarding.

These make the cost associated with our design of including vote links to each vote \textit{negligible.}

\subsection{Optimal Two-Round Voting Process}\label{hotstuffcompare}


HotStuff-2 requires three continuous block intervals to reach a consensus. During these intervals, all honest nodes are aligned, and messages are successfully exchanged before each interval concludes. Similarly, our protocol also necessitates three block intervals: one for proposing the block and two for voting. In the initial block interval (consensus height \( k' \)), block \( B_{k'} \) is proposed and received by all before the interval ends. In the subsequent interval, all honest nodes cast their votes for \( B_{k'} \), and these votes are received by all before the interval's close. By the third interval (vote height \( k'+2 \)), all honest nodes can verify through their vote counts that \( B_{k'} \) has garnered enough support during vote height \( k'+1 \), thereby satisfying Eq.~\ref{acc7} and achieving consensus. Thus, our protocol aligns with the optimal voting rounds of HotStuff-2 and efficiently supports \( O(N) \) view changes when the network is synchronous.

\subsection{Problems Related to Adversarial Leaders}
\subsubsection{Potential DDoS via creating blocks}\label{DDOS}
Based on the rule defined in Sec.~\ref{outputting}, anyone believing they are a leader node can propose a block. Thus, an adversary can send $f$ blocks to the network at each consensus height. But, honest nodes only need to download the block they decide to vote for, and they need to synchronize with other blocks only when switching to branches stemming from those blocks. Ideally, they should synchronize only \emph{one} block per height.
\subsubsection{Sending inconsistent blocks}
Our protocol does not require nodes to align views before voting, potentially a leader node may send inconsistent blocks to other nodes. Such behavior can be detected through vote graphs, and the system can be designed to penalize adversarial leaders later, deterring these types of attacks. If the leader sends $N$ blocks to $N$ nodes, the rounds required for nodes to align remains small, as illustrated by Fig.~\ref{fig:enter-label} with \( m = 100 \). 

\section{Implementation}\label{ae}\vspace{-0.2cm}

\begin{table*}[htbp]
  \centering
  \caption{Experiment Results for good case latency (100 epochs with 1000 transactions emitted every 5 seconds)}\vspace{-0.67cm}
  \label{tab:experiment-results-sync}
  \resizebox{\textwidth}{!}{%
    \begin{tabular}{cc}\makecell{\vspace{-0.25cm}\includegraphics[width=0.3\linewidth]{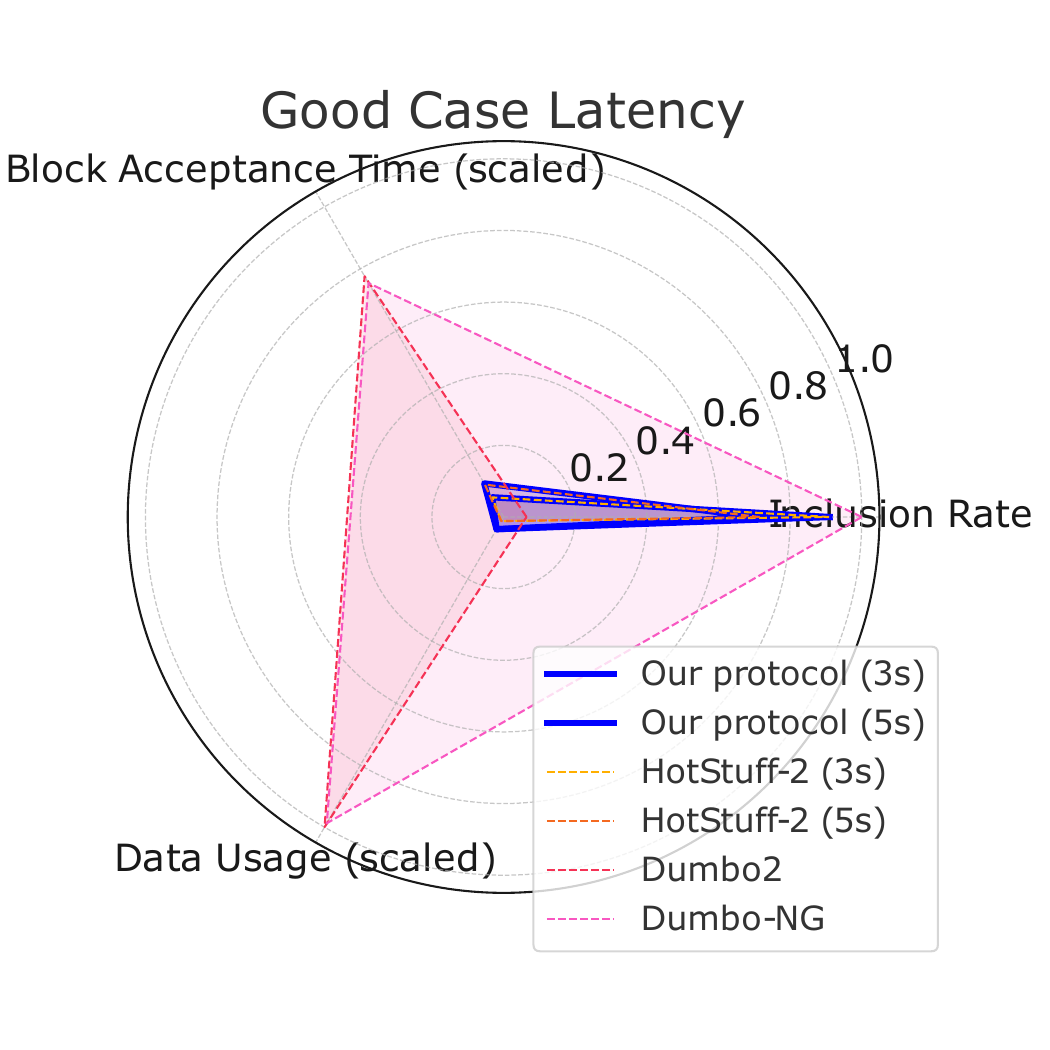}\vspace{-0.5cm}}&
         \begin{tabular}{m{1.9cm}m{2.3cm}m{2.9cm}
    m{3.2cm}m{3cm}}
      \hline
    \textbf{Protocol}& \textbf{Block Interval$\ \ $ (second)}&\textbf{Avg. Transaction Inclusion Rate\footnotemark[1]}
      & \textbf{Avg. Block Acceptance Time (second)\footnotemark[2]}& \textbf{Avg. Data Usage per Node per instance\footnotemark[3]}\\\hline
      \hline
      Our protocol & \centering 3&\centering 0.910 
      & \centering 6.36 &825 KB \\
       HotStuff-2 & \centering 3& \centering 0.909 
       &\centering 6.32 & 278 KB\\
 
     Our protocol & \centering  5 & \centering 0.709 
     &\centering  10.83 &825 KB\\
      HotStuff-2 &\centering  5& \centering 0.715
      &\centering 10.44 & 278 KB\\
                Dumbo2 & \centering Variable&  \centering 0.064
                & \centering 77.62 &20.99 \textbf{MB} \\
      Dumbo-NG & \centering  Variable& \centering 1 
      & \centering 75.41 &20.73 \textbf{MB} \\
      \hline\hline   \end{tabular}\end{tabular}}   
 \resizebox{\textwidth}{!}{%
    \centering
         \begin{tabular}{p{16.2cm}}
 \footnotesize
\footnotemark[1]{The transaction inclusion rate refers to the count of transactions logged to the blockchain when a block is accepted (including the transactions in blocks that are not yet accepted), divided by the total number of transactions ever emitted. For Dumbo, the transaction sets which finished the reliable broadcasting are considered logged to the blockchain.}
\footnotemark[2]{The block acceptance time refers to the time measured from when a block is proposed (sent by the leader node) until it is accepted (when the first node accepts it). For Dumbo, the time starts when the first honest node initiates a new instance (sends its transaction set for this instance) and ends when the first node terminates this instance (committed all valid transaction sets of this instance).}
\footnotemark[3]{The average data usage per node per instance refers to the cost of downloading blocks of a particular height and rounds of votes incurred until the consensus is reached. For example, HotStuff-2 and our approach download only one block except when changing branches, while Dumbo downloads the transaction sets sent by each node. For votes, HotStuff-2 uses a threshold signature to represent the votes of each round, whereas the other approaches download all the votes sent to them. Note that the data for blocks at later heights are not counted. This is reasonable as only the votes contribute to the acceptance of the block, but later blocks do not.}
      \end{tabular}
  }\vspace{-0.2cm}
\end{table*}

Our implementation served two main purposes: (1) demonstrating that our protocol achieved consensus using the same voting rounds as HotStuff-2 in good-case latency settings; and (2) highlighting the robustness of our protocol in bad-case latency settings, where it significantly outperformed asynchronous blockchains in terms of transaction inclusion, steady blockchain progression, and data usage. 

We did not experiment with DAG-based protocols. They exhibited similar complexity to ACS-based protocols. Comparisons with DAG-based protocols are given in Appx.~\ref{DAG_}.

\noindent \textbf{Setup:} The experiments were executed on two servers, each featuring a 56-core Intel(R) Xeon(R) Gold 5120 CPU running at 2.20GHz, complemented by 500GB of memory. There were 100 nodes in total for each approach, 33 of them were adversarial. As discussed in Sec.~\ref{sig}, placing Dumbo2 and Dumbo-NG in a non-fully-meshed authenticated network would have incurred significant costs. Thus, in the experiment, all nodes were fully meshed. To mimic geographic distribution, message delays between sender and receiver were varying between 80 to 290 milliseconds.

Our protocol, Dumbo2~\cite{guo2020dumbo}, Dumbo-NG\cite{gao2022dumbo}, and the partially synchronous HotStuff-2~\cite{yin2019hotstuff} were tested in our experiment. For Dumbo, we set the message size ($|M|$) to $2.05\times 10^{6}$ bits and the security parameter $\lambda=256$.

At the start of each 5 seconds, 1000 identical transactions (each sized 256 bytes) in total appeared in each node's local database. This was regardless of the progress of the blockchains. The leader node grouped the first 1000 transactions not written to the blockchain to form a block. For nodes in Dumbo, they broadcasted the uncommitted transactions, which together sized $2.05\times 10^{6}$ bits. Further justifications of the experiment metrics are given in Appx.~\ref{REM}.

\subsection{Good-case Latency}
In this experiment, each honest message was received before 290 milliseconds after it was sent.
\subsubsection{Methodology}
We conducted an experiment of 100 blockchain epochs for each approach. We set $\Delta BI=3$ and $\Delta BI=5$, so that the leader node would generate a new block after $3$ or $5$ seconds once it received and verified the block of the latest height. The block interval for Hotstuff was also set to $3$ and $5$, respectively. For fair comparison with Hotstuff, we used the same design that the leader node gathered the votes (no need for exactly $N-f$ votes, see the amendment in Sec.~\ref{relaxing}) and sent with its block.

Regarding voting, adversarial nodes in our algorithm were constrained to make compliant votes. If they voted, they expedited consensus. We, therefore, assumed that adversarial nodes in our protocol and Hotstuff would attempt to create forks when selected as a leader, and otherwise kept silent, akin to the behavior of silent Byzantine nodes in Dumbo.

We tested the speed (seconds) for a block to be accepted in consensus for different approaches. We also tested the transaction inclusion rate when a block of a block height was accepted in consensus. The transaction inclusion rate was defined as $\frac{TIB}{OT}$, where $TIB$ was the transaction logged to the blockchain (including the blocks not yet accepted), and $OT$ was the overall transaction emitted to the network. We also tested the overall data usage per instance.

\subsubsection{Results}\textit{Our protocol used a similar average block acceptance time compared to HotStuff-2, while HotStuff-2 required smaller data; both approaches largely outperformed Dumbo2 and Dumbo-NG due to reduced complexity, as expected. Tbl.~\ref{tab:experiment-results-sync} reported the experiment results.}

\noindent \textbf{Data usage and block acceptance time:} Compared to Dumbo2 and Dumbo-NG, where each node received $N-f$ sets (each of size $|M|$) of transactions, and they were all overlapping transactions, the size amounted to $16.37$ MB already, not to consider other size associated. In contrast, our protocol and HotStuff-2 received only one block every 5 seconds at most. Compared to HotStuff-2, our protocol had increased data usage because it included links in the votes and necessitated indicating others' votes from the preceding voting round, leading to higher data demands.

The time for our protocol and HotStuff-2 to reach consensus depended on the pre-set block interval as that determined when the next leader would propose a block and a new voting round would not start until the next block was received. Dumbo2 did not have the adversarial leader problem, but it had to wait for one instance to finish before starting the next block height, resulting in a block interval of approximately 77.6 seconds on average. Dumbo-NG ran $N$ continuous broadcasting instances, allowing broadcasting phases of future consensus-derivation instances to start early.

\noindent \textbf{Transaction inclusion rate:} For HotStuff-2 and our protocol, each time an adversarial leader was selected, it resulted in a waste of 3 or 5 seconds due to forking or remaining silent. In such cases, this adversarial leader effectively did not include any new transactions in the blockchain because the next leader had to repeat the job of this adversarial leader. The likelihood of adversarial nodes being selected was $\frac{f}{N} \approx \frac{1}{3}$. This explained the transaction inclusion rate and partly contributed to the average block acceptance time. In the 3-second version, we generated a new block faster than the system emitted new transactions, thus achieving a higher inclusion rate. But, in the 5-second version, both approaches may have had an inclusion rate only closer to $\frac{2}{3}$ because only about $\frac{2}{3}$ of the time, an honest leader was selected.

\subsection{Bad-case Latency in Asynchronous Network}\label{exa}
The delay of an honest message was arbitrary but finite. To imitate this condition, we set each message from the slowest honest node to arrive at each node exactly at the end of every ten seconds. However, the nodes were unaware of the latest time when a message would be received.

\subsubsection{Methodology}
As in the good-case latency network, we conducted experiments over 100 blockchain epochs for each approach. HotStuff could not function in an asynchronous network and was excluded from this experiment.

\begin{table*}[htbp]
  \centering
  \caption{Experiment Results for bad case latency (100 epochs with 1000 transactions emitted every 5 seconds. Each node will receive only one message from the slowest honest node every 10 seconds)}\vspace{-0.2cm}
  \label{tab:async-consensus-results}
  \vspace{-.5cm}\resizebox{\textwidth}{!}{%
  \vspace{-.4cm}
  \begin{tabular}{cc}
      \makecell{\includegraphics[width=0.262\linewidth]{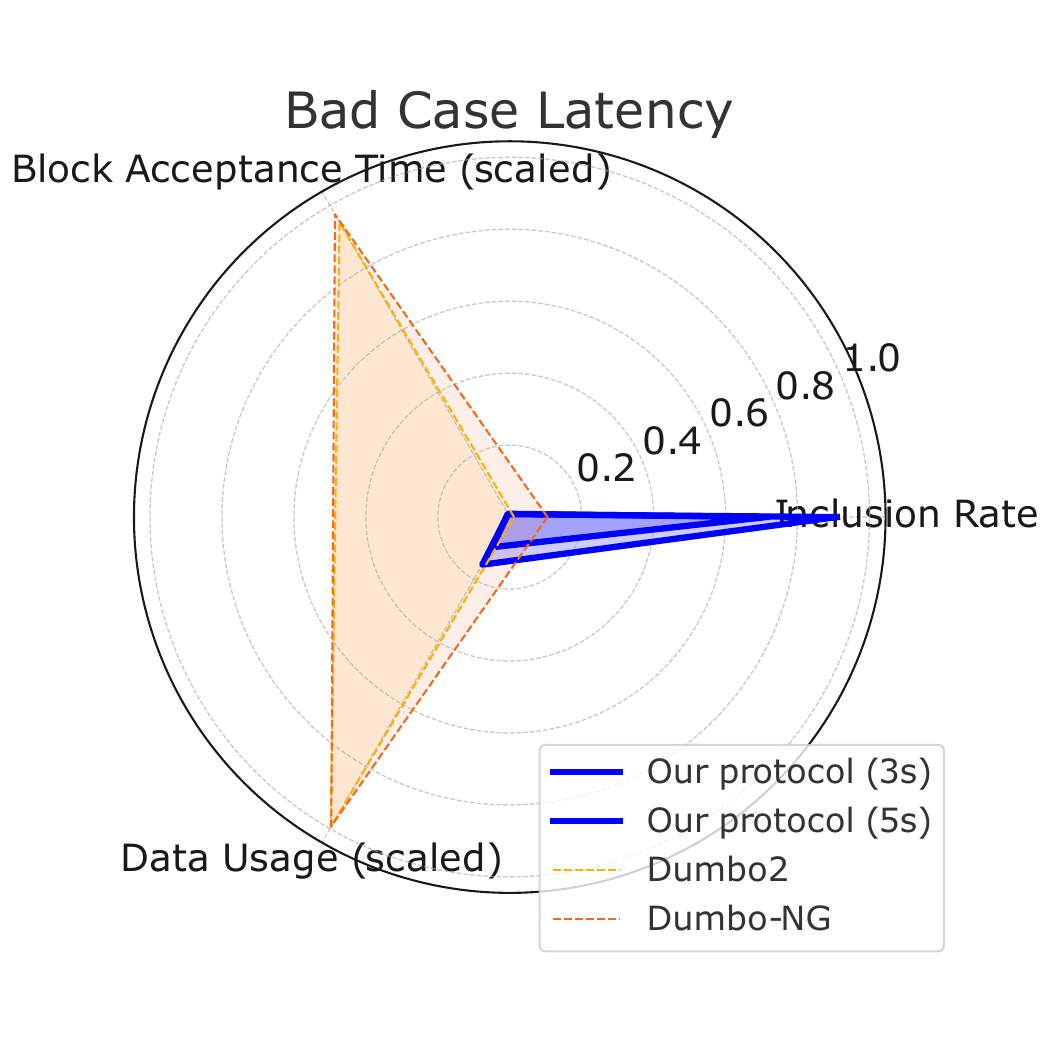}}&
     \begin{tabular}{m{2cm}m{2.4cm}m{2.9cm}
         m{3.3cm}m{3.3cm}}
      \hline
      \textbf{Protocol} &   \textbf{Block Interval $\ \ $ (second)}&\textbf{Avg. Transaction Inclusion Rate}
      & \textbf{Avg. Block Acceptance Time (second)}& \textbf{Avg. Data Usage per Node per instance\footnotemark[1]}\\\hline
      \hline
      Our protocol &\centering 3& \centering 0.909
      &20.68 &3.21 \textbf{MB} \\
       Our protocol &\centering 5& \centering 0.713 
       &23.42 &2.03 \textbf{MB}\\
          Dumbo2 &\centering Variable&\centering 0.012 
          & 2081.42& 20.90 \textbf{MB} \\
      Dumbo-NG &\centering Variable&  \centering 0.105 
      & 2138.01& 20.84 \textbf{MB}  \\
   
      \hline\hline
    \end{tabular}%
     \end{tabular}
  }\vspace{-.5cm}
  \begin{tabular}{p{17cm}}\footnotesize
      \footnotemark[1] Directly comparing the Avg. Data Usage per Node per instance favors Dumbo as the instances in our protocol are nested together, meaning that the votes of a vote height are also counted in the data usage of other instances. In contrast, the votes of Dumbo are only for a particular instance.\vspace{-1cm}
       \end{tabular}
\end{table*}

\subsubsection{Results} \textit{As expected, our protocol largely outperformed Dumbo2 and Dumbo-NG due to reduced complexity.} Tbl.~\ref{tab:async-consensus-results} presented the experimental results. The transaction inclusion rate for our protocol was similar in both good and bad-case latency because the only factors that affected it in our experiment were the rate of adversarial leaders and the pre-set block interval. The average data usage per node per instance increased because rounds of voting occurred while awaiting the votes from the slowest honest node. Note that it was \emph{unfair} to directly compare our protocol with Dumbo in terms of the average data usage per node per instance because the votes of a vote height counted in our protocol were also for other instances, and therefore, also counted in the other instances. Consequently, our data usage should be considered much smaller than Dumbo2 and Dumbo-NG.

Our protocol, with reduced complexity compared to Dumbo, achieves much faster block acceptance under the same slowest honest node. The complexity of Dumbo impacts its block interval and transaction inclusion rate in this experiment. For example, $N$ parallel Consistent Broadcasts end sequentially because the slowest honest node only replies to one instance every 10 seconds (only one of its messages is received by others). Dumbo-NG used more time because some slow honest messages are used for the broadcasting instances for later epochs, which started before the current epoch ended.

\subsection{Short Block Intervals}
The previous results typically showed that most nodes communicated effectively before the block interval ended, with at most a 10-second delay for only one honest node. Consequently, instances where honest nodes voted for different blocks at each consensus height were uncommon.

\begin{figure}[htbp]
  \centering
  \includegraphics[width=.5\textwidth]{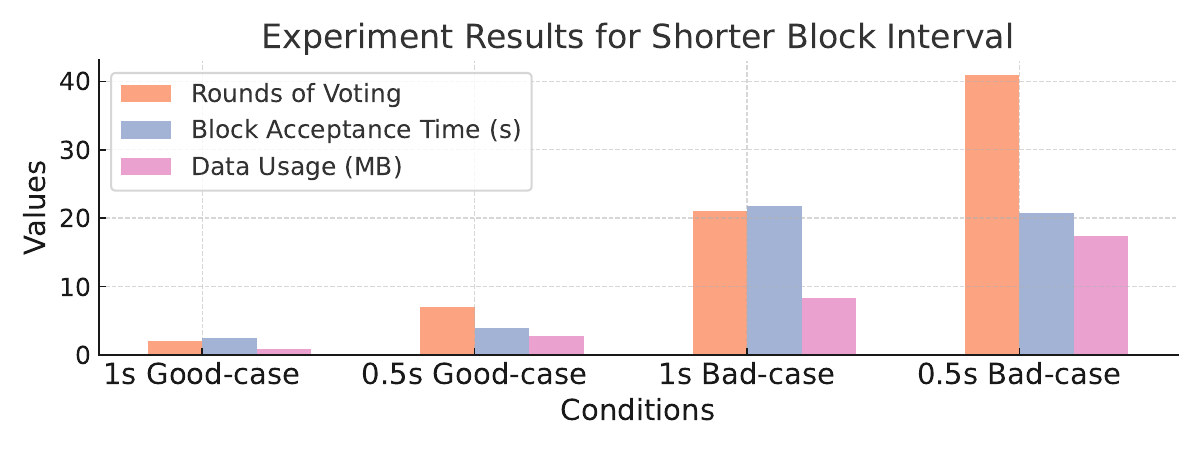}
  \caption{Experiment Results for shorter block interval (100 epochs with 1000 transactions emitted every 5 secs). }
  \label{tab:async-multi-results1}
\end{figure}

We evaluated the effects of shorter block intervals---1 and 0.5 seconds---on node synchronization (see Fig.~\ref{tab:async-multi-results1}), using the same experimental setup as given in previous experiments. The findings indicated that it took 2 and 7 rounds to reach consensus under good-case latency, whereas under bad-case latency, it took 21 and 41 rounds, respectively. Here, the extended voting period for bad-case latency primarily served to wait for the slowest honest node's vote. Notably, while nodes aligned to the same branch within a few rounds, reaching a consensus required receiving the votes from the slowest honest nodes.

The average transaction inclusion rate was 100\% because transactions were emitted only every 5 seconds, and either 5 or 10 consensus heights had passed, which was sufficient for including these transactions. We did not encounter the situation of five or ten consecutive adversarial leaders in our experiment. We wanted to emphasize again that the average data usage per node per instance did not provide a good evaluation of the data usage because the instances were nested together, and the votes counted for the data usage of one instance were also counted in other instances. If we had not implemented the relaxing design in Sec.~\ref{relaxing}, and instead asked nodes to vote only when $N-f$ votes from the previous height had been received, then we would have had fewer rounds of voting and less data usage per node per instance in the bad-case latency scenarios.

We did not test the case where the adversary constantly controlled entire network delays to such an extent that it could precisely and constantly decide which $N-f$ votes from the last height would be received first by each node. Even when using the leader-collecting system, where the leader could decide which votes to send, it would require the honest nodes always to recognize the Byzantine node as the leader node to perform this attack, which was impossible. The amendment to overcome this improbable attack was outlined in Sec.~\ref{aced}. We believed this attack abused the asynchronous network assumption and was not realistic, especially in the decentralized or at least federated networks.

\section{Conclusion}
Our robust validated strong BFT consensus protocol is specially designed for leader-and-vote-based blockchains, particularly suited for asynchronous networks. It surpasses existing asynchronous blockchain solutions in terms of operational simplicity and consensus efficiency. With adversarial tolerance comparable to that of binary byzantine agreement algorithms, it ensures the seamless progression of a blockchain without requiring immediate consistency among nodes. By addressing challenges in achieving termination and agreement properties within asynchronous environments, our protocol ensures liveness in consensus and significantly reduces message complexity associated with asynchronous BFT algorithms.

{\bibliographystyle{acm}
\bibliography{main}}

\appendix
\section{Related Works}\label{new33}

\subsection{Strong BFT Consensus Protocols}\label{strong}
Strong BFT consensus \cite{10.1145/872035.872066,neiger1994distributed,berman1989asymptotically} is a variant of vote-based BFT consensus where there are more than one input value for nodes to vote on, and multiple voting rounds are used for the honest nodes to decide on the same input value eventually. Each node votes for an input value in each round, but it can vote for different input values in different rounds. 

The protocol instructs the nodes to vote for an input value at random using a (multi-valued) shared random coin \cite{neiger1994distributed} from all remaining input values at each round, except for the first round, where the nodes vote for the input value of their choice. When node $A$ receives at least $N-f$ votes in a voting round, it undergoes three synchronization phases:

1). Node $A$ sends a report to all nodes with the IDs of the input values that have received more than $f$ votes.

2). Once node $A$ receives at least $N-f$ reports, if $N-f$ reports contain the ID of some input values, then these input values are considered `confirmed.' Subsequently, $A$ sends a confirmation message containing the IDs of the `confirmed' input values to all nodes. 

3). When node $A$ has received $N-f$ confirmations, if more than $f$ nodes have `confirmed' an input value, then the input value is `accepted' into the next round of voting. Note that this input value does not necessarily have to be confirmed by $A$ in the second phase. All input values that were not `accepted' are eliminated.

Afterwards, $A$ enters a new voting round with the remaining input values until only one input value remains.

\noindent\textbf{Safety.} \label{strongsafety}A Byzantine node may double vote or send its votes only to a subset of honest nodes. However, this cannot trick some nodes into believing that an input value has been advanced to a new round when it actually did not. This is because less than $N-f$ nodes will confirm that the input value has received more than $f$ votes. There could be different sets of $N-f$ reports, resulting in some nodes considering that an input value was not confirmed, even though it was confirmed. But, these nodes will work this out eventually during the third phase as there must be more than $f$ nodes who have accepted this input value into the next round. So, after each voting round, the honest nodes reach the same set of remaining input values.

The security lower bound for the asynchronous version is $f \leq \lfloor (N-1)/ (m + 1)\rfloor$ and the synchronous version $f \leq  \lfloor (N-1)/m)\rfloor $\cite{10.1145/872035.872066,neiger1994distributed}, where $m>2$ is the maximum number of input values allowed. But, in Sec.~\ref{VSC}, we argue that this is not the case when 
implemented in blockchains. 

\noindent\textbf{Liveness.} An adversary cannot prevent the reduction of input values because it only controls $f$ votes. An input value that is not voted on by any honest node will not receive more than $f$ votes and will not advance to the next round. 


\noindent\textbf{Message count and message complexity.} Each node sends three messages per voting round: one about which input value it votes for, one about which input values have received more than $f$ votes to its knowledge, and one indicates the input values for the next round. Thus, there are $3N$  messages, each sent to all $N$ nodes. 
The  message count and message complexity are $O(m+3\theta N)$ and $O(mN+3\theta N^2)$, respectively.

\subsubsection{Challenges in guaranteeing liveness of consensus derivation with the leader-based design}\label{CHAA}
The strong BFT consensus protocol in Appx.~\ref{strong} requires a known set of $m$ input values and tolerates only a small $f$. It does not consider the origin of the input values. If $m$ leaders propose them, then consensus liveness in asynchronous settings is not  guaranteed unless $m \geq N-f$. Also, when $m > 2$ and \(f \leq \lfloor (N-1)/3 \rfloor\), it is possible that no input value receives more than \(f\) votes, making it impossible to discard a block that doesn't receive $f+1$ votes.

In contrast, our protocol does not require a known $m$ but ensures that, via multiple voting rounds, honest nodes will converge to the same value. 

\subsection{ACS-based Protocols}
HoneyBadgerBFT~\cite{miller2016honey} is the first asynchronous blockchain protocol. Each consensus derivation process first uses $N$ instances of reliable broadcast to establish the sets of transactions submitted by each node; then it instantiates $N$ instances of ABA (Asynchronous Byzantine Agreement). The communication cost ACS instantiation in HoneyBadgerBFT is $O(N^2|v|+\lambda N^3\log N)$ bits, assuming $|v|$ is the largest size of any node's input and $\lambda$ is a security parameter. Dumbo1~\cite{guo2020dumbo} optimizes HoneyBadgerBFT by running only $k$ instead of $N$ ABA instances, where $k\leq N$ is a security parameter. Dumbo2~\cite{guo2020dumbo} uses MVBA instead of ABA. The total communication cost of both Dumbo1 and Dumbo2 is $O(N^2|M|+\lambda N^3\log N)$ bits, where $|M|$ represents the size of MVBA’s input values \cite{guo2020dumbo}. Overall, it requires $O(N^3)$ messages. Dumbo-NG~\cite{gao2022dumbo} further optimizes Dumbo2 by minimizing the message complexity to $O(N^2)$ using a protocol called sDumbo-DL and solving the transaction inclusion blockage by running $N$ ever-running broadcasts and a sequence of Byzantine Agreements.  However, the overall communication cost remains unchanged.  Other works \cite{blum2023abraxas, duan2022waterbear} either optimize communication cost or have hybrid designs adapting to various networks.

\subsection{DAG-based Protocols}  \label{DAG_}
Direct Acyclic Graph (DAG)-based BFT protocols \cite{keidar2021all,danezis2022narwhal,spiegelman2022bullshark} use DAGs to transmit transactions. Each node sends a block linked to $N-f$ blocks from the preceding height. A leader block is periodically selected to order and finalize the linked blocks. To avoid equivocation, they must reliably broadcast each block, driving the overall complexity to $O(N^3)$.  
A promising recent preprint~\cite{jovanovic2024mahi} aims to reach consensus while allowing equivocation in asynchronous networks during the process, thereby reducing the overall complexity.  

DAG-based protocols are particularly useful for avoiding delays caused by slow or faulty leaders. Instead of selecting a leader first and waiting for the leader node to propose a block of transactions for each block height, every node proposes a batch of transactions it knows, with the leader node selected later and the batch of the leader node be finalized as the consensus. This approach prevents delays caused by a pre-determined leader node but comes with the trade-off of increased data usage. Since each block is linked to $N-f$ blocks from the last round, each block can also be viewed as a vote for the previous rounds.  

Our protocol is very similar to DAG-based protocols if all $N$ nodes propose a block at each block height. In this setting, our protocol could also be used to avoid the delays caused by faulty or slow leader nodes. However, there are significant differences even in this setting:  
\begin{enumerate}
    \item The method of achieving consensus is entirely different. DAG-based protocols still require a deterministic leader, albeit with the leader election postponed until after nodes broadcast their blocks. To ensure a deterministic leader, they must determine the exact blocks broadcast to avoid equivocation. Optimizations so far only seek to patch equivocation issues without involving Byzantine reliable broadcasting. In contrast, our approach does not require a deterministic leader at all. Instead, we leverage the vote graph to gradually force all nodes to align with the same branch.
    \item Since votes and blocks are separated in our approach, we require all nodes to wait until they receive $N-f$ votes instead of $N-f$ blocks. The size of blocks is significantly larger than the size of votes. Thus, our protocol should be much faster than DAG-based protocols and demand less data.  
    \item Nodes in our protocol do not need to download all $N-f$ blocks per round for the protocol to function. Instead, each node only needs to download the chain of blocks leading up to the block it votes for. As such, unless it needs to switch to another branch, each node downloads only one block per height in normal cases, regardless of how many other blocks are present. In Appx.~\ref{aLCR}, we show that even when all nodes propose a block, nodes converge in only few rounds. Thus, nodes only switch branches few times before a block is consensually accepted. This supports our claim of significantly reduced data usage and significantly less message complexity compared to DAG-based protocols.
\end{enumerate}  

Thus, despite also using graphs, our approach is much more similar to classical leader-based methods such as PBFT and HotStuff where nodes only download a block that they vote for, making our protocol distinct from current DAG-based approaches while requiring much less data.

\section{Proofs}

\printProofs

\section{Further Analysis and Experiment}
\subsection{Remarks on the FLP Result}

The Fischer-Lynch-Paterson (FLP) result~\cite{fischer1985impossibility} demonstrates that achieving agreement in an asynchronous message-passing system is impossible if even a single crash failure is allowed, unless the basic model is augmented with additional mechanisms, such as randomization or failure detectors.

Our protocol addresses this limitation through two key mechanisms:  
\begin{enumerate}
    \item \textbf{Randomness from network delays:} The inherent unpredictability of network delays allows nodes to construct different vote graphs randomly, influencing their voting decisions in subsequent rounds.
    \item \textbf{Common coin flip:} In cases where the randomness of network delays is insufficient or absent, a common coin flip is employed as a source of randomization.
\end{enumerate}

These features ensure that our protocol incorporates sufficient randomization to circumvent the constraints imposed by the FLP result. Consequently, the correctness of our protocol is not affected by the FLP impossibility.

\subsection{Remarks on Consensus Chain Propagation}\label{C3}
In Sec.~\ref{outputting}, we assumed that nodes have access to a local clock. However, this assumption does not imply that the system is partially synchronous or synchronous in terms of communication. The local clock is neither established through communication among nodes nor required to align perfectly. The sole purpose of this assumption is to provide a trigger for certain nodes to act as leader nodes, proposing blocks periodically.

Since the uniqueness of the leader node is not required, this mechanism can be achieved through other means of randomness. For example, a node could monitor its own CPU speed, which is ostensibly unpredictable. When the CPU speed coincidentally matches a pre-set random number, the node proposes a new block. This approach also helps ensure liveness by facilitating the extension of the consensus chain.

\subsection{Remarks on Vote Compliance Checking}\label{C2}
It is well known that a perfect Byzantine failure detector does not exist~\cite{cachin2011reliable, dubois2015weakest}. The core issue lies in the fact that, without guaranteeing that all honest nodes observe the same information, they cannot make consistent judgments. Indeed, in asynchronous networks, there is no assurance that honest nodes will observe the same information, not even eventually, as nodes may send different messages to different recipients.

However, our protocol focuses on detecting non-compliant votes rather than Byzantine nodes. Specifically, nodes are required to justify their vote for a particular block by demonstrating that this block belongs to the most supported branch according to the vote graph derived from the vote. When each node receives a vote, they can construct identical vote graphs using this vote, making it straightforward for all honest nodes to reach the same judgment based on the information in the vote graph.

Furthermore, as each vote graph is constructed using the $N-f$ vote graphs from the previous height---many of which originate from honest nodes---the adversary's voting decisions are progressively constrained round by round.

Thus, our protocol does not aim to detect Byzantine nodes directly. Instead, it systematically reduces the valid choices for nodes in each round through vote compliance checking.

\subsection{Remarks on Experimental Metrics}\label{REM}
Our experimental metrics differ from those used in existing asynchronous BFT consensus protocols, which measure transactions accepted per second (TPS) with varying batch/block sizes. Directly relating the batch size of an ACS-based protocol to TPS suggests a hidden assumption that transactions among all batches are distinct. This is unrealistic in a real-world blockchain where users may submit their transactions to multiple nodes in an attempt to have their transactions included (because they do not know if a particular node is honest) in the blockchain. Our experimental setup, thus, considers a realistic scenario in which 1,000 transactions are emitted to all nodes every 5 seconds, causing large overlap in the batches sent by each node in ACS-based protocols.

In our protocol, block proposals are entirely decoupled from the voting process, allowing new blocks to be broadcast at regular, fixed intervals, regardless of whether the previous block has been accepted. Variations in block intervals can affect the number of branches in the network. Because leader nodes extend the consensus chain known to themselves, different consensus chains with a common prefix can develop into branches. If new blocks are proposed at a very fast rate, it will result in a great number of branches before nodes convergence into the same one. This scenario provides an interesting experimental setup for evaluating performance across different block intervals and other factors that might influence branch formation.

The batch/block size significantly influences latency (the time taken to accept transactions) in ACS-based or DAG-based protocols. This distinction arises because, in those protocols, every node must broadcast its set of transactions and each node must receive at least $N-f$ batches of transactions per instance; thus, small changes in batch size can drastically impact overall data usage, affecting both the number of transactions per batch and the latency.

In contrast, in our protocol, only the leader node broadcasts a block under normal circumstances. Ideally, a node only downloads $N$ votes and a block per height. A node is required to download other blocks only when it switches to vote in different branches. Even if adversaries broadcast more blocks and create additional branches, each node only synchronizes the chain of blocks leading up to the block it votes for when switching branches. Consequently, the batch/block size has rather minimal impact on experimental outcomes. Instead, the frequency with which nodes switch to vote for different branches affects performance. 

Thus, experimenting with different batch/block sizes becomes uninteresting and not very insightful in terms of TPS variations. With that said, in our experiment, $TPS = \frac{\text{Transactions in a block}}{\text{Avg. Block Acceptance Time}}$.

\subsection{Multiple Concurrent Leader Nodes}\label{multi}

The transaction inclusion blockages may occur if the volume of uncommitted transactions exceeds the capacity of the system. For example, when $\Delta BI=5$ and new transactions of a block size are emitted to each node also in 5 seconds, as set up in Sec.~\ref{ae}, the transactions are delayed in being included in the blockchain due to a faulty leader wasting 5 seconds of its leader time. As such, only $2/3$ of transactions are included in the blockchain because only $2/3$ of the leaders are honest.

To investigate the role of strong BFT consensus in mitigating transaction inclusion blockages, we explore the allowance of multiple concurrent leader nodes to propose blocks in parallel. Specifically, using the consensus chain propagation method discussed in Sec.~\ref{outputting}, for three concurrent leaders, $i=\lfloor\frac{T}{\Delta BI}\rfloor \mod N$, nodes $i$, $i+1$ and $i+2$ each propose a block for every period $\Delta BI$.  The blockchain progression experiences slowdowns only when all leaders within the slots are adversarial. HotStuff-2 cannot support this design because the honest votes' view regarding the block with the largest block height would be divided among multiple blocks. As a result, multiple highest chains can exist simultaneously, and the block cannot be finalized.

Tbl.~\ref{tab:async-multi-results} shows our experiment results using the same experimental setup as given in Sec.~\ref{ae}. For two or three concurrent leaders, the chance of wasting time due to forking is ${(\frac{f}{N})}^2$ and ${(\frac{f}{N})}^3$ respectively, with an increase in acceptance time and data usage per instance due to more rounds of voting required. We only test up to three concurrent leaders because $97\%$ of average transaction inclusion rate is already good enough and the number of voting rounds used with a larger number of leaders can be seen from Fig.~\ref{fig:enter-label}.

\begin{table}[htbp!]
  \centering
  \caption{Experiment results for multiple concurrent leaders (100 epochs with 1000 transactions emitted every 5 secs). }
  \label{tab:async-multi-results}
  \resizebox{0.5\textwidth}{!}{%
  
       \begin{tabular}{|p{2.9cm}|p{1.4cm}p{1.4cm}p{1.2cm}p{1.2cm}|}
      \hline
      \textbf{} & \textbf{Two leaders} & \textbf{Three leaders} & \textbf{Two leaders} & \textbf{Three leaders} \\
      \hline\hline
      \textbf{Latency} & Good-case & Good-case & Bad-case & Bad-case \\
      \hline
      \textbf{Block Interval (second)} & 5 & 5 & 5 & 5 \\
      \hline
      \textbf{Avg. Transaction Inclusion Rate} & 0.9094 & 0.9731 & 0.9104 & 0.9698 \\
      \hline
      \textbf{Avg. Block Acceptance Time (second)} & 13.7 & 15.68 & 30.35 & 40.55 \\
      \hline
      \textbf{Avg. Data Usage per Node per instance (MB)} & 1.12 & 1.45 & 2.16 & 2.96 \\
      \hline\hline
    \end{tabular}%
  }
\end{table}

\section{Rounds Until Convergence}\label{aLCR}
\begin{figure}[h!]
    \centering
\includegraphics[width=0.5\textwidth]{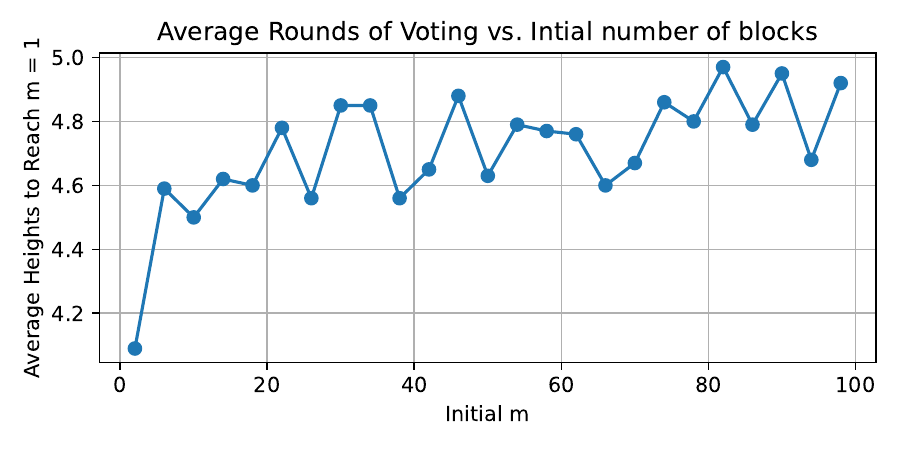}\vspace{-0.5cm}
    \caption{The relationship between the initial number of blocks and the average voting rounds used until convergence over 100 simulations for each initial m.}

    \label{fig:enter-label}
\end{figure}

We present straightforward simulations demonstrating nodes arriving at a common branch. As detailed in the paper, nodes must contribute their votes to the branch with the highest support as suggested by the vote count ($\operatorname{VC}$) of the votes, otherwise the votes are non-compliant votes and are not counted. We did simple simulations using a Python code provided in Code Listing 1 to test the average number of voting rounds (vote heights) used until all nodes converge on the same branch in practice.  In each run of the simulation, each node votes for a random block among the $m$ blocks at the first round of voting. 
To simulate the random network delays in asynchronous network, each node will receive a random $N-f$ compliant set of votes. They then vote for the most supported one determined through a connected vote graph constructed from this random set of $N-f$ compliant votes. This simulation faithfully represents the normal scenarios where the adversary lacks control over the network delays. The Byzantine nodes must also vote in the same way, as otherwise, their votes are not compliant votes. 
Fig.~\ref{fig:enter-label} shows the simulation results for a \(N=100\), \(f=33\) system with varying initial numbers of blocks ($m$). Each simulation with an initial \(m\) is repeated 100 times. Fig.~\ref{fig:enter-label} suggests that even when all nodes generated a block (initial \(m=100\)), it still took no more than five vote heights for all nodes to reach the same branch.

Moreover, we illustrate four simulation outcomes in Fig.~\ref{fig:_}, Fig.~\ref{fig:_1}, Fig.~\ref{fig:_2}, and Fig.~\ref{fig:_3}, considering initial scenarios where initial \(m\) is either 100 or 2, and blocks receive votes randomly or uniformly at the outset. In these visualizations, the array `Votes $\rightarrow$ []' signifies the votes at a specific height, where the `i-th' node cast a vote for a block indexed by the value in the `i'-th slot of the array. This array sizes $N$. As evidenced by the outcomes, all simulations converged within 4 rounds.

\begin{lstlisting}[language=Python, caption=Simulation of voting process,
    backgroundcolor=\color{backcolour},   
    commentstyle=\color{codegreen},
    keywordstyle=\color{magenta},
    numberstyle=\tiny\color{codegray},
    stringstyle=\color{codepurple},
    basicstyle=\ttfamily,
    breakatwhitespace=false,         
    breaklines=true,                 
    captionpos=b,                    
    keepspaces=true,                             
    showspaces=false,                
    showstringspaces=false,
]
import random
# Parameters
N = 100  # Total number of nodes
f = 33   # Number of faulty nodes
num_blocks = 100  # Number of block candidates (m)
heights = 100  # Number of heights to simulate
assert N == 3 * f + 1
# Initialize structures to hold vote data and observed votes
votes_at_height = [[[] for _ in range(N)] for _ in range(heights)]
observed_votes = [[set() for _ in range(N)] for _ in range(heights)]
current_votes = [[[] for _ in range(N)] for _ in range(heights)]
current_votes [0] =[random.randint(1, num_blocks) for _ in range(N)] # Initial votes, randomly distributed 
#current_votes [0] = [i % num_blocks + 1 for i in range(N)]  # Initial votes, evenly distributed
def update_observed_indices(node, height, visible_indices):
    """Updates the record of seen votes at a particular height."""
    if height==-1:
        return 
    observed_votes[height][node]=observed_votes[height][node].union(visible_indices)
    for i in visible_indices:
        update_observed_indices(node, height-1, observed_votes[height-1][i])
def calculate_VC_i(height):
    """Calculate the Vote Count for each node at a given height."""
    VC = [{block: 0 for block in range(1, num_blocks + 1)} for _ in range(N)]
    for node in range(N):
        for h in range(height+1):
            for vote in votes_at_height[h][node]:
                VC[node][vote] += 1
    return VC
def simulate_voting_round(height):
    """Simulate a voting round where each node updates its vote based on observed and potentially supported votes."""
    global current_votes
    new_votes = []
    for node in range(N):
        visible_indices = random.sample(range(N), N - f)
        update_observed_indices(node, height, visible_indices)
        height_=height
        while height_>=0:
            votes_at_height[height_][node]  = [current_votes[height_][i] for i in observed_votes[height_][node]]
            height_-=1
        VC = calculate_VC_i(height)
        vote_counts = {block: VC[node][block] for block in range(1, num_blocks + 1)}
        max_support = max(vote_counts.values())
        candidates = [block for block, count in vote_counts.items() if count == max_support]
        chosen_block = random.choice(candidates)
        new_votes.append(chosen_block)
    current_votes[height+1] = new_votes
def alignment_reached(h):
    """Check if all nodes have the same vote."""
    return len(set(current_votes[h+1])) == 1
# Main simulation loop
print(f"Height {1}: Votes -> {current_votes[0]}")
for height in range(heights):
    simulate_voting_round(height)
    print(f"Height {height+2}: Votes -> {current_votes[height+1]}")
    if alignment_reached(height):
        print(f"Convergence reached at height {height+2}.")
        break
\end{lstlisting}

\begin{figure}[h!]
\begin{lstlisting}[backgroundcolor=\color{mygray},   
    commentstyle=\color{codegreen},
    keywordstyle=\color{magenta},
    numberstyle=\tiny\color{codegray},
    stringstyle=\color{codepurple},
    basicstyle=\ttfamily,
    breakatwhitespace=false,         
    breaklines=true,                 
    captionpos=b,                    
    keepspaces=true,                 
    showspaces=false,                
    showstringspaces=false,
    showtabs=false,                  
    tabsize=1]
Height 1: Votes -> [64, 75, 63, 74, 57, 45, 31, 11, 2, 9, 76, 46, 41, 23, 78, 75, 99, 86, 89, 13, 35, 59, 27, 56, 54, 91, 41, 29, 15, 35, 12, 70, 55, 79, 72, 60, 21, 11, 53, 95, 53, 31, 74, 90, 93, 62, 32, 99, 12, 61, 48, 29, 74, 69, 8, 19, 48, 52, 68, 18, 84, 44, 80, 91, 9, 47, 77, 68, 45, 16, 64, 12, 79, 95, 98, 13, 87, 42, 8, 71, 8, 12, 51, 85, 31, 58, 46, 60, 85, 94, 38, 5, 34, 42, 79, 21, 73, 19, 17, 54]
Height 2: Votes -> [12, 31, 12, 12, 74, 31, 12, 12, 12, 12, 79, 12, 31, 12, 74, 68, 74, 79, 79, 12, 8, 8, 8, 12, 12, 8, 79, 12, 12, 79, 12, 8, 12, 12, 12, 8, 12, 31, 12, 79, 8, 74, 8, 79, 79, 79, 12, 31, 74, 12, 31, 12, 74, 12, 12, 54, 12, 79, 74, 79, 79, 12, 12, 74, 74, 12, 12, 12, 42, 79, 12, 8, 12, 79, 74, 12, 31, 8, 8, 31, 19, 46, 79, 12, 12, 79, 31, 74, 12, 74, 12, 12, 31, 12, 74, 12, 12, 12, 12, 12]
Height 3: Votes -> [12, 12, 12, 12, 12, 12, 12, 12, 12, 12, 12, 12, 12, 12, 12, 12, 12, 12, 12, 12, 12, 12, 12, 12, 12, 12, 12, 12, 12, 12, 12, 12, 12, 12, 12, 12, 12, 12, 12, 12, 12, 12, 12, 12, 12, 12, 12, 12, 12, 12, 12, 12, 12, 12, 12, 12, 12, 12, 12, 12, 12, 12, 12, 12, 12, 12, 12, 12, 12, 12, 12, 12, 12, 12, 12, 12, 12, 12, 12, 12, 12, 12, 12, 12, 12, 12, 12, 12, 12, 12, 12, 12, 12, 12, 12, 12, 12, 12, 12, 12]
Convergence reached at height 3.
\end{lstlisting}
\caption{The simulation results for N=100, f=33, initial m=100, blocks received random votes at the first vote height}
    \label{fig:_}
\end{figure}

\begin{figure}[h!]
\includegraphics[width=0.5\textwidth]{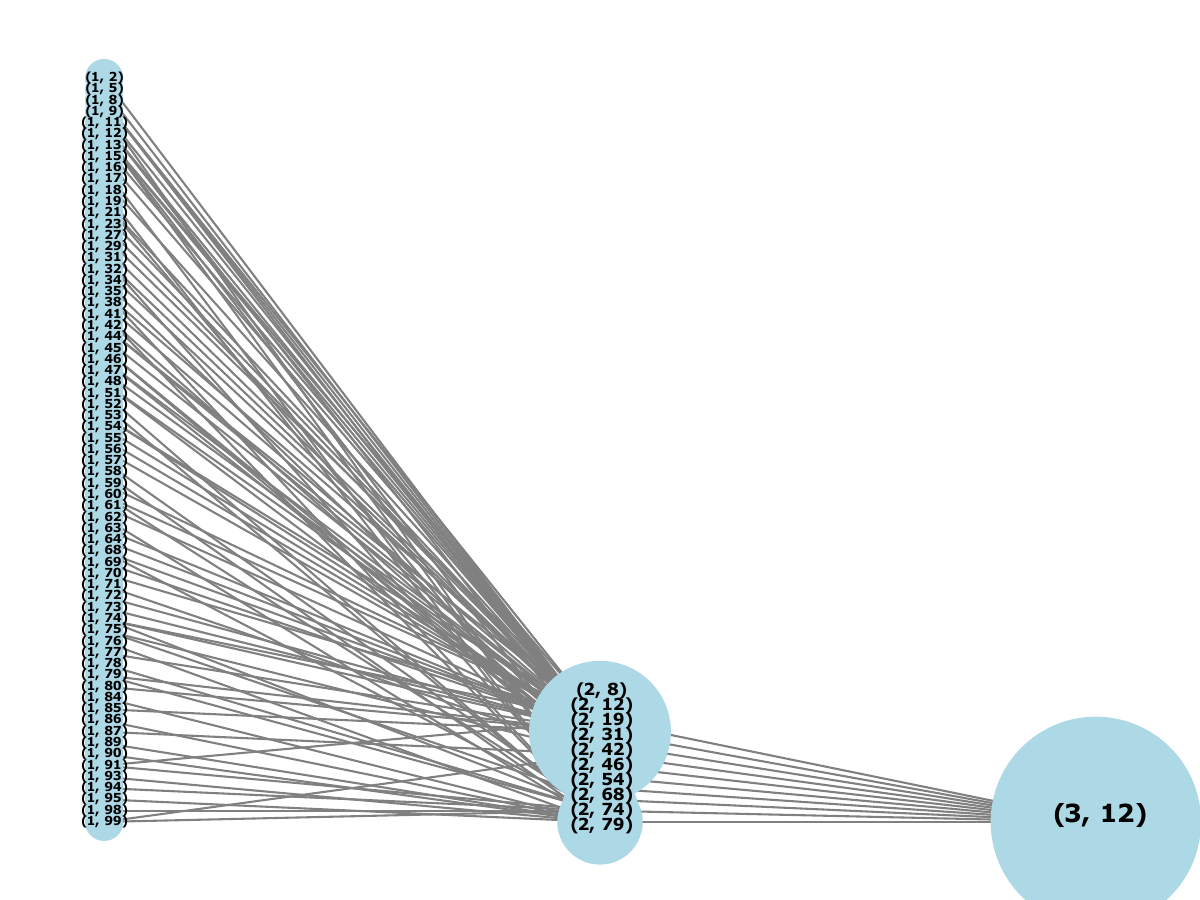}
\caption{The visualization of Fig.~\ref{fig:_}. $(X,Y)$ refers to the vote for block ID $Y$ at height $X$. Convergence reached at height 3. }
    \label{fig:_1_}
\end{figure}
\begin{figure}[h!]
\begin{lstlisting}[backgroundcolor=\color{mygray},   
    commentstyle=\color{codegreen},
    keywordstyle=\color{magenta},
    numberstyle=\tiny\color{codegray},
    stringstyle=\color{codepurple},
    basicstyle=\ttfamily,
    breakatwhitespace=false,         
    breaklines=true,                 
    captionpos=b,                    
    keepspaces=true,                 
    numbers=left,                    
    numbersep=1pt,                  
    showspaces=false,                
    showstringspaces=false,
    showtabs=false,                  
    tabsize=1]
Height 1: Votes -> [1, 2, 3, 4, 5, 6, 7, 8, 9, 10, 11, 12, 13, 14, 15, 16, 17, 18, 19, 20, 21, 22, 23, 24, 25, 26, 27, 28, 29, 30, 31, 32, 33, 34, 35, 36, 37, 38, 39, 40, 41, 42, 43, 44, 45, 46, 47, 48, 49, 50, 51, 52, 53, 54, 55, 56, 57, 58, 59, 60, 61, 62, 63, 64, 65, 66, 67, 68, 69, 70, 71, 72, 73, 74, 75, 76, 77, 78, 79, 80, 81, 82, 83, 84, 85, 86, 87, 88, 89, 90, 91, 92, 93, 94, 95, 96, 97, 98, 99, 100]
Height 2: Votes -> [97, 1, 7, 44, 19, 57, 3, 89, 14, 82, 13, 23, 56, 91, 88, 34, 31, 92, 46, 19, 42, 89, 26, 19, 6, 29, 28, 52, 74, 89, 28, 71, 48, 86, 84, 51, 50, 74, 94, 91, 93, 58, 8, 40, 29, 99, 16, 67, 33, 30, 97, 24, 77, 79, 64, 32, 16, 4, 32, 28, 18, 19, 76, 27, 2, 51, 68, 27, 28, 65, 59, 19, 50, 91, 17, 1, 78, 29, 64, 3, 81, 51, 17, 84, 51, 96, 5, 4, 44, 94, 18, 73, 47, 60, 39, 42, 89, 24, 7, 9]
Height 3: Votes -> [19, 19, 28, 51, 28, 19, 19, 51, 19, 19, 19, 29, 89, 19, 89, 51, 28, 51, 19, 19, 51, 19, 89, 91, 51, 51, 19, 28, 89, 19, 29, 89, 28, 28, 19, 51, 89, 28, 28, 51, 19, 19, 19, 19, 89, 51, 28, 51, 19, 19, 89, 89, 19, 89, 89, 89, 28, 51, 28, 19, 19, 19, 28, 28, 19, 19, 89, 89, 19, 19, 89, 28, 51, 89, 19, 19, 28, 51, 89, 28, 51, 51, 89, 89, 51, 29, 19, 19, 51, 19, 89, 19, 19, 28, 89, 89, 51, 89, 28, 19]
Height 4: Votes -> [19, 19, 19, 19, 19, 19, 19, 19, 19, 19, 19, 19, 19, 19, 19, 19, 19, 19, 19, 19, 19, 19, 19, 19, 19, 19, 19, 19, 19, 19, 19, 19, 19, 19, 19, 19, 19, 19, 19, 19, 19, 19, 19, 19, 19, 19, 19, 19, 19, 19, 19, 19, 19, 19, 19, 19, 19, 19, 19, 19, 19, 19, 19, 19, 19, 19, 19, 19, 19, 19, 19, 19, 19, 19, 19, 19, 19, 19, 19, 19, 19, 19, 19, 19, 19, 19, 19, 19, 19, 19, 19, 19, 19, 19, 19, 19, 19, 19, 19, 19]
Convergence reached at height 4.
\end{lstlisting}
\caption{The simulation results for N=100, f=33, initial m=100, blocks received the same votes at the first vote height}
    \label{fig:_1}
\end{figure}
\begin{figure}[h!]
\includegraphics[width=0.5\textwidth]{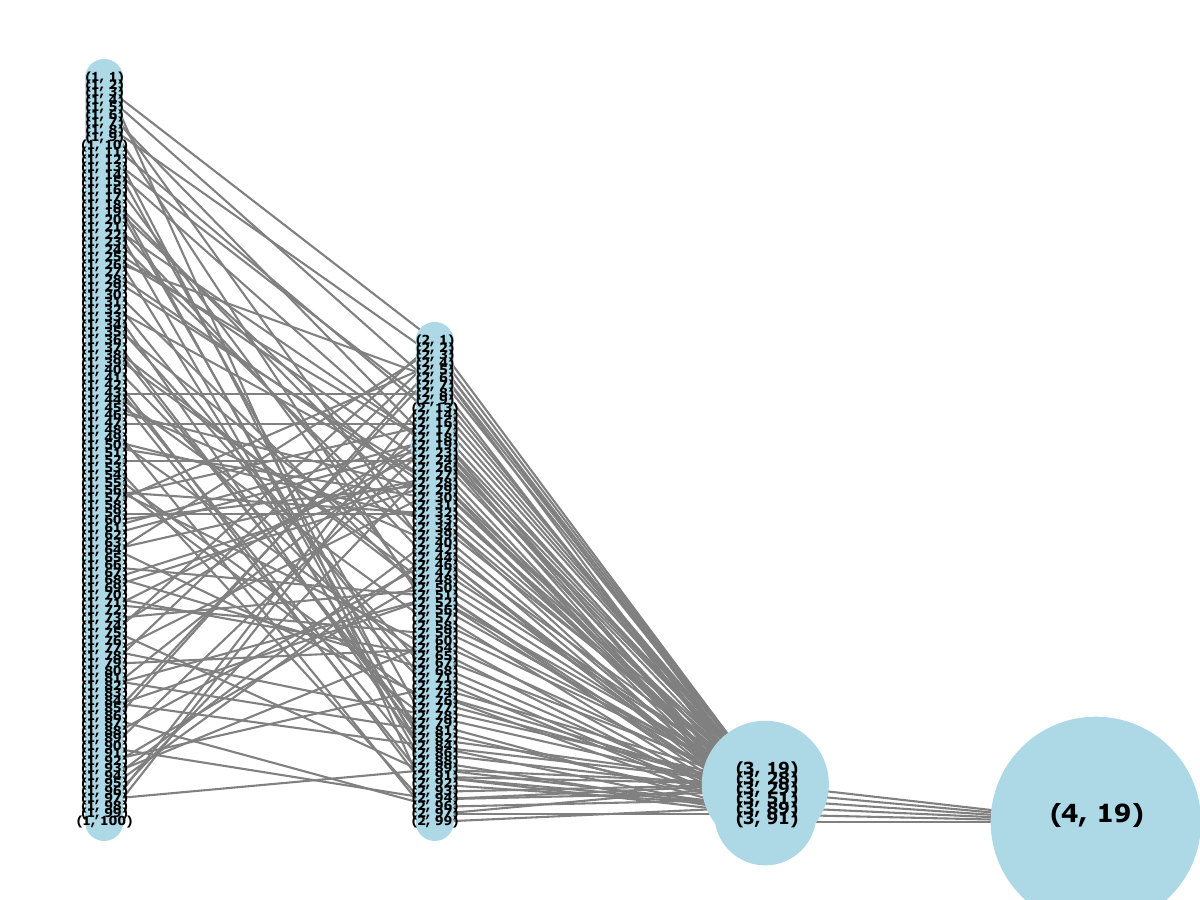}
\caption{The visualization of Fig.~\ref{fig:_1}. $(X,Y)$ refers to the vote for block ID $Y$ at height $X$. Convergence reached at height 4}
    \label{fig:_1_1}
\end{figure}
\begin{figure}[h!]
\begin{lstlisting}[backgroundcolor=\color{mygray},   
    commentstyle=\color{codegreen},
    keywordstyle=\color{magenta},
    numberstyle=\tiny\color{codegray},
    stringstyle=\color{codepurple},
    basicstyle=\ttfamily,
    breakatwhitespace=false,         
    breaklines=true,                 
    captionpos=b,                    
    keepspaces=true,                 
    numbers=left,                    
    numbersep=1pt,                  
    showspaces=false,                
    showstringspaces=false,
    showtabs=false,                  
    tabsize=1]
Height 1: Votes -> [1, 2, 1, 2, 1, 2, 1, 2, 1, 2, 1, 2, 1, 2, 1, 2, 1, 2, 1, 2, 1, 2, 1, 2, 1, 2, 1, 2, 1, 2, 1, 2, 1, 2, 1, 2, 1, 2, 1, 2, 1, 2, 1, 2, 1, 2, 1, 2, 1, 2, 1, 2, 1, 2, 1, 2, 1, 2, 1, 2, 1, 2, 1, 2, 1, 2, 1, 2, 1, 2, 1, 2, 1, 2, 1, 2, 1, 2, 1, 2, 1, 2, 1, 2, 1, 2, 1, 2, 1, 2, 1, 2, 1, 2, 1, 2, 1, 2, 1, 2]
Height 2: Votes -> [2, 2, 1, 2, 2, 1, 2, 2, 1, 1, 2, 2, 1, 2, 2, 2, 1, 1, 2, 1, 1, 2, 2, 1, 2, 1, 2, 1, 1, 2, 1, 2, 1, 2, 2, 1, 2, 2, 2, 1, 1, 1, 1, 2, 2, 2, 2, 1, 2, 2, 2, 2, 1, 2, 2, 1, 1, 2, 2, 2, 1, 2, 2, 2, 2, 2, 1, 1, 1, 1, 1, 2, 1, 2, 2, 1, 2, 2, 2, 1, 1, 2, 1, 1, 2, 2, 2, 2, 1, 2, 1, 1, 1, 2, 2, 1, 2, 1, 2, 1]
Height 3: Votes -> [2, 2, 2, 2, 2, 2, 2, 2, 2, 2, 2, 2, 2, 2, 2, 2, 2, 2, 2, 2, 2, 2, 2, 2, 1, 2, 2, 2, 2, 2, 2, 2, 2, 2, 2, 2, 2, 2, 2, 2, 2, 2, 2, 2, 2, 2, 2, 2, 2, 2, 2, 2, 2, 2, 2, 2, 2, 2, 2, 2, 2, 2, 2, 2, 2, 1, 2, 2, 2, 2, 2, 2, 2, 2, 2, 2, 2, 2, 2, 2, 2, 2, 2, 2, 2, 2, 2, 2, 2, 2, 2, 2, 2, 2, 2, 2, 2, 2, 2, 2]
Height 4: Votes -> [2, 2, 2, 2, 2, 2, 2, 2, 2, 2, 2, 2, 2, 2, 2, 2, 2, 2, 2, 2, 2, 2, 2, 2, 2, 2, 2, 2, 2, 2, 2, 2, 2, 2, 2, 2, 2, 2, 2, 2, 2, 2, 2, 2, 2, 2, 2, 2, 2, 2, 2, 2, 2, 2, 2, 2, 2, 2, 2, 2, 2, 2, 2, 2, 2, 2, 2, 2, 2, 2, 2, 2, 2, 2, 2, 2, 2, 2, 2, 2, 2, 2, 2, 2, 2, 2, 2, 2, 2, 2, 2, 2, 2, 2, 2, 2, 2, 2, 2, 2]
Convergence reached at height 4.
\end{lstlisting}
\caption{The simulation results for N=100, f=33, initial m=2, blocks received the same votes at the first vote height}
    \label{fig:_2}
\end{figure}
\begin{figure}[h!]
\includegraphics[width=0.5\textwidth]{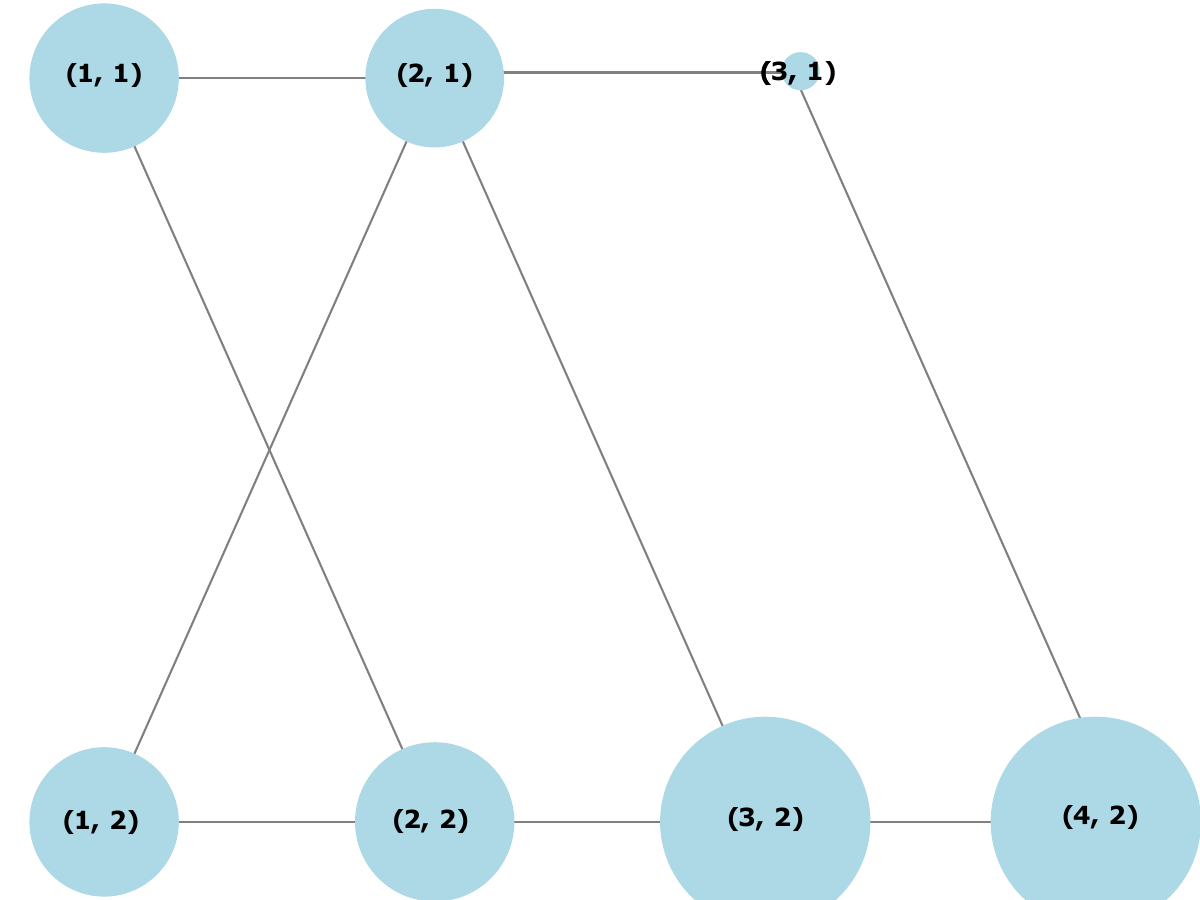}
\caption{The visualization of Fig.~\ref{fig:_2}. $(X,Y)$ refers to the vote for block ID $Y$ at height $X$. Convergence reached at height 4}
    \label{fig:_2_1}
\end{figure}
\begin{figure}[h!]
\begin{lstlisting}[backgroundcolor=\color{mygray},   
    commentstyle=\color{codegreen},
    keywordstyle=\color{magenta},
    numberstyle=\tiny\color{codegray},
    stringstyle=\color{codepurple},
    basicstyle=\ttfamily,
    breakatwhitespace=false,         
    breaklines=true,                 
    captionpos=b,                    
    keepspaces=true,                 
    numbers=left,                    
    numbersep=1pt,                  
    showspaces=false,                
    showstringspaces=false,
    showtabs=false,                  
    tabsize=1]
Height 1: Votes -> [1, 2, 2, 2, 1, 2, 2, 1, 1, 2, 1, 2, 1, 2, 2, 1, 1, 2, 1, 2, 1, 2, 1, 2, 2, 2, 2, 1, 2, 2, 1, 2, 2, 2, 1, 2, 2, 1, 2, 2, 1, 1, 2, 2, 2, 1, 2, 2, 1, 1, 1, 2, 1, 1, 1, 2, 2, 2, 1, 2, 2, 1, 1, 2, 2, 2, 2, 2, 2, 1, 2, 2, 2, 1, 2, 1, 2, 2, 1, 2, 2, 1, 2, 1, 1, 2, 1, 1, 2, 2, 1, 2, 1, 1, 1, 2, 1, 1, 2, 1]
Height 2: Votes -> [2, 2, 2, 2, 2, 2, 2, 2, 2, 2, 2, 2, 2, 2, 2, 1, 2, 2, 2, 2, 2, 2, 2, 2, 2, 2, 2, 2, 2, 2, 2, 2, 2, 2, 2, 2, 2, 2, 2, 2, 2, 2, 2, 2, 2, 2, 2, 2, 2, 2, 2, 2, 2, 2, 2, 2, 2, 2, 2, 1, 2, 2, 2, 2, 2, 2, 2, 2, 2, 2, 2, 2, 2, 2, 2, 2, 2, 2, 2, 2, 2, 2, 2, 2, 2, 2, 2, 2, 2, 1, 2, 2, 2, 2, 2, 2, 2, 2, 2, 2]
Height 3: Votes -> [2, 2, 2, 2, 2, 2, 2, 2, 2, 2, 2, 2, 2, 2, 2, 2, 2, 2, 2, 2, 2, 2, 2, 2, 2, 2, 2, 2, 2, 2, 2, 2, 2, 2, 2, 2, 2, 2, 2, 2, 2, 2, 2, 2, 2, 2, 2, 2, 2, 2, 2, 2, 2, 2, 2, 2, 2, 2, 2, 2, 2, 2, 2, 2, 2, 2, 2, 2, 2, 2, 2, 2, 2, 2, 2, 2, 2, 2, 2, 2, 2, 2, 2, 2, 2, 2, 2, 2, 2, 2, 2, 2, 2, 2, 2, 2, 2, 2, 2, 2]
Convergence reached at height 3.
\end{lstlisting}
\caption{The simulation results for N=100, f=33, initial m=2, blocks received random votes at the first vote height}
    \label{fig:_3}
\end{figure}
\begin{figure}[h!]
\includegraphics[width=0.5\textwidth]{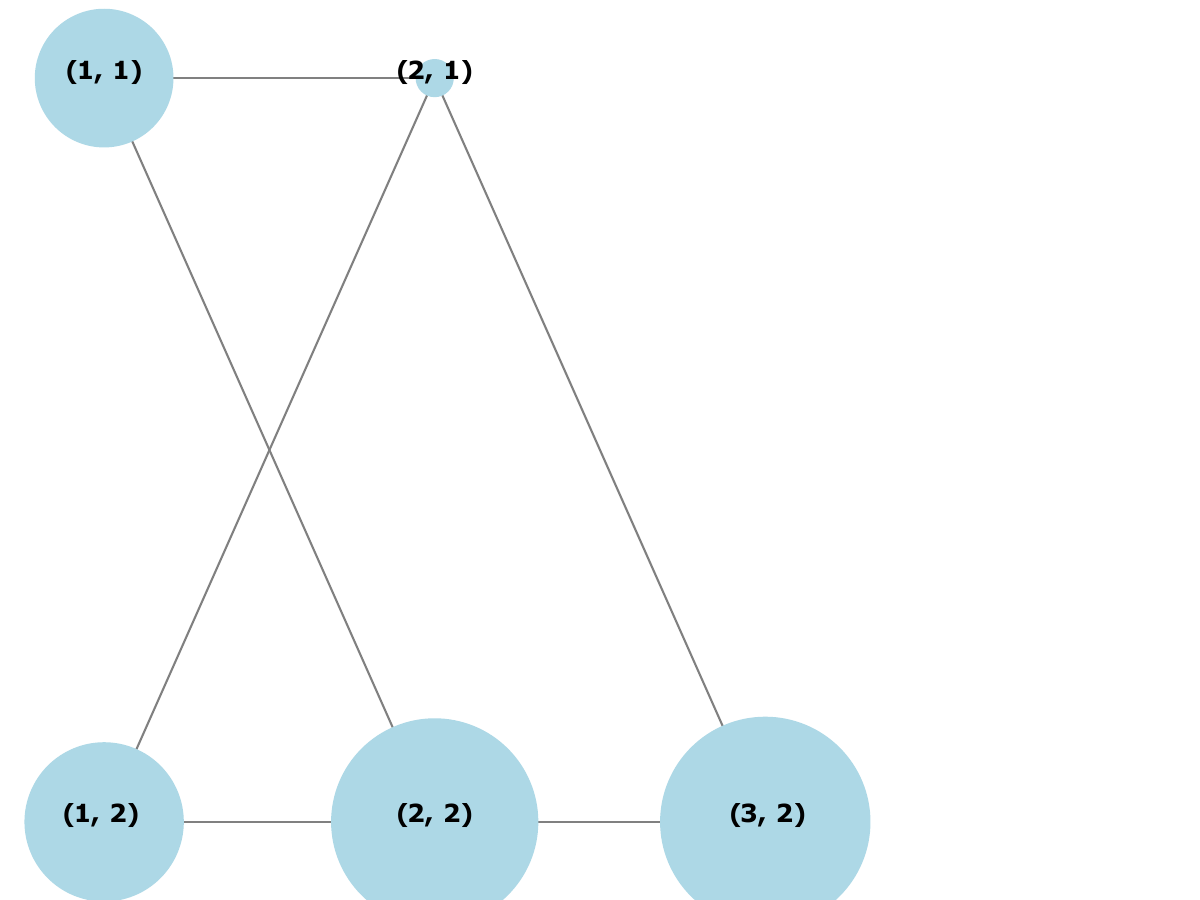}
\caption{The visualization of Fig.~\ref{fig:_3}. $(X,Y)$ refers to the vote for block ID $Y$ at height $X$. Convergence reached at height 3}
    \label{fig:_3_1}
\end{figure}

\section{Using the Validated Strong BFT Consensus Visualization Tool}
\label{G}
To help understand our Validated Strong BFT Consensus protocol, we provide a \emph{toy} experiment as well as its visualization tool for illustration, which visualizes the protocol in action, demonstrating how BFT consensus is achieved in our protocol in an asynchronous environment.

The interactive version of our tool can be accessed via the following anonymous link: \url{https://waterandcola.github.io/validated-strong-consensus/index.html}

\textbf{Steps to Try the Tool:} 

1. \textbf{Access the Webpage:} Navigate to the provided URL to open the interactive tool in your browser. No local setup or additional dependencies are needed.

2. \textbf{Replay the Consensus Process:} The example provided simulates a consensus network of 7 nodes ($N=7$) with a fault tolerance of up to 2 faulty nodes ($f=2$). Each node proposes blocks and casts votes in asynchronous rounds to reach consensus. In this process, all 7 nodes behave normally. This is because if the adversary cast non-compliant votes, their votes are not processed, resulting in 5 active honest nodes and two silent Byzantine nodes. Since each node only votes after receiving $N-f$ votes from the previous height, having only 5 active honest nodes leads to very fast branch alignment (since each one will have the same view after one voting round), making it less interesting for illustration.

 3. \textbf{View Logs and Replay Consensus:}
    \begin{itemize}
        \item You can either use the pre-generated logs already displayed in the tool or \emph{create your own logs} using a small real experiment via \noindent \url {https://github.com/WaterandCola/validated-strong-consensus/blob/main/vote.py}. 
        
        This python code runs seven networked peers running validated strong BFT consensus protocol.
        \item The logs show details such as vote heights, block IDs, and the vote graph $G$ and the vote count $\operatorname{VC}$ for each vote, helping you understand how nodes make decisions and how consensus emerges.
        \item Details see \url {https://github.com/WaterandCola/validated-strong-consensus/blob/main/README.md}
        
    \end{itemize}

4. \textbf{Visualize the Consensus Graph:}
    \begin{itemize}
        \item The tool generates a dynamic graph where \emph{blocks} (squares) and \emph{votes} (circles) are displayed. 
        \item \emph{Green blocks} represent blocks that have reached consensus after accumulating the required number of votes.
        \item You can hover over votes to see additional information about the vote, voter graph, vote counter and the blocks they support, as shown in Fig.~\ref{fig:-label}.
        \begin{figure*}[h!]
    \centering
    \includegraphics[width=\linewidth]{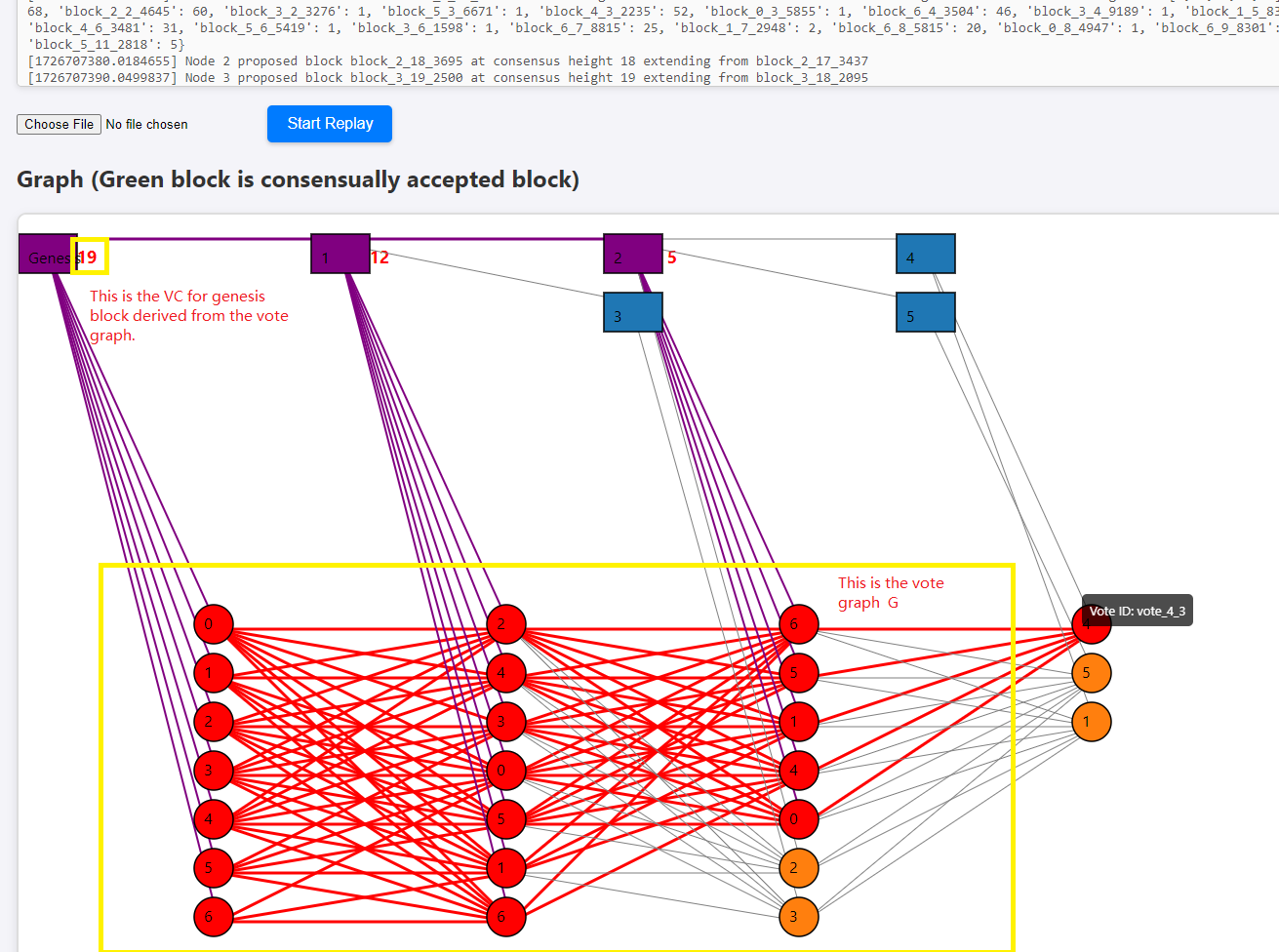}
    \caption{An explanation for visualization tool (\url{https://waterandcola.github.io/validated-strong-consensus/index.html}).}
    \label{fig:-label}
\end{figure*}
    \end{itemize}

\end{document}